\documentclass[12pt,a4paper]{article}
\usepackage{amssymb}

\usepackage{amsmath}
\usepackage{latexsym}

\usepackage[dvips]{graphicx} 

\numberwithin{equation}{section}
\newcommand{\RPtwo}{\mathbb{RP}^2}

\providecommand{\abs}[1]{\lvert#1\rvert}

\newcommand{\lanln}[1]{$\langle$\texttt{arXiv:#1}$\rangle$}

\newtheorem{proposition}{Proposition}[section]

\begin{document}

\title{Geons with spin and charge}
\author{Jorma Louko$^1$\thanks{%
jorma.louko@nottingham.ac.uk}, \ Robert B. Mann$^{2,3}$\thanks{%
mann@avatar.uwaterloo.ca} \ and Donald Marolf$\,{}^4$\thanks{%
marolf@physics.ucsb.edu} \\
\noalign{\vspace{3ex}} $^1${\small \textit{School of Mathematical Sciences,
University of Nottingham, Nottingham NG7 2RD, UK} }\\
\noalign{\smallskip} $^2${\small \textit{Perimeter Institute for Theoretical
Physics, Ontario N2J 2W9, Canada} }\\
\noalign{\smallskip} $^3${\small \textit{Department of Physics, University
of Waterloo Waterloo, Ontario N2L 3G1, Canada} }\\
\noalign{\smallskip} $^4${\small \textit{Physics Department, UCSB, Santa
Barbara, CA 93106, USA} }\\
\noalign{\vspace{3ex}}\\
{\small {(Revised February 2005)} }\\
\noalign{\vspace{2ex}} 
{\small {$\langle$\texttt{arXiv:gr-qc/0412012}$\rangle$} }}
\date{}
\maketitle

\begin{abstract}
We construct new geon-type black holes in $D\geq 4$ dimensions for
Einstein's theory coupled to gauge fields. A~static nondegenerate vacuum
black hole has a geon quotient provided the spatial section admits a
suitable discrete isometry, and an antisymmetric tensor field of rank $2$ or 
$D-2$ with a pure $\boldsymbol{F}^{2}$ action can 
be included by
an appropriate 
(and in most cases nontrivial) choice of the 
field strength bundle.
We find rotating geons as quotients of the Myers-Perry(-AdS) solution when 
$D$ is odd and not equal to~$7$. For other $D$ we show that such rotating
geons, if they exist at all, cannot be continuously deformed to zero angular
momentum. With a negative cosmological constant, we construct geons with
angular momenta on a torus at the infinity. As an example of a nonabelian
gauge field, we show that the $D=4$ spherically symmetric $\mathrm{SU}(2)$
black hole admits a geon version with a trivial gauge bundle. Various
generalisations, including both black-brane geons and Yang-Mills theories
with Chern-Simons terms, are briefly discussed.
\end{abstract}

\tableofcontents


\section{Introduction}

The term ``geon'', short for ``gravitational-electromagnetic entity'',
was introduced in 1955 by John Archibald Wheeler to denote a classical
gravitational configuration, possibly coupled to electromagnetism or
other zero-mass fields, that appears as a long-lived massive object
when observed from a distance but is not a black hole
\cite{wheeler-geon,brill-wheeler-geon,ernst-geon-var,ernst-geon,%
MW-geondata,brill-hartle-geon}.  
While a precise definition of a geon
seems not to have been sought, the examples studied in
\cite{wheeler-geon,brill-wheeler-geon,ernst-geon-var,ernst-geon,%
MW-geondata,brill-hartle-geon} 
had spatial topology $\mathbb{R}^3$ and
with one exception an asymptotically flat infinity, where mass could
be defined by what are now known as ADM methods. The one exception was
Melvin's magnetic universe
\cite{ernst-geon,melvin-pl}, which was subsequently deemed not to qualify as a
geon, precisely because of its cylindrical infinity~\cite{melvin-prd}.
Unfortunately, while Melvin's universe is stable
\cite{melvin-prd,thorne-Cen,thorne-melvinstab}, 
these geons were not, owing to the tendency of a massless field either
to disperse to infinity or to collapse into a black hole. This
tendency has more recently been much studied in the context of
critical phenomena in gravitational collapse~\cite{gundlach-rev}.

In 1985, Sorkin \cite{sorkin-topogeon} denoted by ``topological
geons'' gravitational configurations whose spatial geometry is
asymptotically flat and has the topology of a compact manifold with
one puncture, the omitted point being at the asymptotic infinity. This
generalises Wheeler's geon in the sense of allowing nontrivial
spatial topology and a black hole horizon. An example of a topological
geon that is a black hole is the space and time orientable
$\mathbb{Z}_{2}$ quotient of the Kruskal manifold known as the
$\mathbb{RP}^3$ geon~\cite{FriedSchWi}. This spacetime 
has an asymptotically flat exterior region isometric to
a standard Schwarzschild exterior and is hence an eternal black hole. 
However, while in Kruskal the
spatial hypersurfaces are $\mathbb{R}
\times S^2$ wormholes with two asymptopias, the spatial topology of
the $\mathbb{RP}^{3}$ geon is
$\mathbb{RP}^{3}\setminus\{$point at infinity$\}$.\footnote{The
time-symmetric initial data for the $\mathbb{RP}^{3}$ geon was
discussed
prior to \cite{FriedSchWi} 
in \cite{MW-geondata}
and \cite{giulini-thesis,giulini-multiRP3s}. An early discussion on
quotients of Kruskal can be found in \cite{szekeres} and a modern one
in~\cite{chamblin-gibbons}. The Euclidean-signature section of the
$\mathbb{RP}^{3}$ geon is discussed in~\cite{louko-marolf-rp3}.}

As an eternal black and white hole spacetime, the $\mathbb{RP}^{3}$
geon is not an object one would expect to be formed in astrophysical
processes. Its interest is that as an \emph{un\/}conventional black
hole, it provides an arena for probing our understanding of black
holes both in the classical and quantum contexts. For example, in
\cite{FriedSchWi} the $\mathbb{RP}^{3}$ geon was used to illustrate a
classical topological censorship theorem, and the Hamiltonian dynamics
of spherically symmetric spacetimes with geon-type boundary conditions
was investigated in \cite{louko-whiting,friedman-louko-winters}. The
Hawking(-Unruh) effect on the $\mathbb{RP}^{3}$ geon was analysed for
scalar and spinor fields in \cite{louko-marolf-rp3,langlois-rp3},
addressing questions about the geon's entropy and its statistical
mechanical interpretation. The $(2+1)$ dimensional asymptotically AdS
analogue of the $\mathbb{RP}^{3}$ geon was used in
\cite{louko-marolf,lou-ma-ross,malda-eternal} to probe AdS/CFT
correspondence. 

The purpose of this paper is to present new families of geon-type black
holes in $D\geq 4$ spacetime dimensions for Einstein's theory coupled to
gauge fields. We shall not attempt to give a precise definition of
``geon-type black hole'', but we require the spacetimes to be time
orientable and foliated by spacelike hypersurfaces with a single asymptotic
region. We also require the asymptotic region to be stationary and `simple'
in a sense that allows conserved charges to be defined by appropriate
integrals. We also consider geon variants of certain black branes that
appear in string theory. We find that while the $\mathbb{RP}^{3}$ geon
generalises readily into static vacuum geons in any dimension, the inclusion
of $\mathrm{U}(1)$ gauge fields is more subtle and depends sensitively on
the dimension, in most cases requiring 
the field strength to be a section of a bundle 
that is twisted by the spatial fundamental group or by some quotient
thereof. Such twisting is in particular required for electrically charged
Reissner-Nordstr\"{o}m{} in any spacetime dimension and for magnetically
charged Reissner-Nordstr\"{o}m{} in even spacetime dimensions. Similarly,
while the rotating $(2+1)$-dimensional black hole has geon-type variants for
any values of the angular momentum \cite%
{aminneborg-bengtsson-holst,brill-samos,brill-weimar}, we find that that the
options for building geons from the rotating Kerr-Myers-Perry(-AdS)
solutions \cite%
{myers-perry,hawking-hunter-tr,gibbons-lu-page-pope1,%
gibbons-lu-page-pope2,gibbons-perry-pope}
depend sensitively on the dimension. With a negative cosmological constant,
we also present rotating geons with a flat horizon. As an example of a
nonabelian gauge field, we show that the 
$D=4$ spherically symmetric $\mathrm{SU}(2)$ black hole 
\cite{bizon,kuenzle-masood,win-stability,bjoraker-hoso-small,%
bjoraker-hoso-big,win-sar-even,win-sar-odd}
admits a geon variant with a trivial gauge bundle.

We use metric signature $(-++\cdots )$. Static Einstein-$\mathrm{U}(1)$
geons are discussed in section \ref{sec:static} and geons with angular
momentum in section~\ref{sec:angmom}. The $D=4$ spherically symmetric $%
\mathrm{SU}(2)$ geon is constructed in section~\ref{sec:su2geon}. Section %
\ref{sec:discussion} presents a summary and concluding remarks.

\section{Static Einstein-$\mathrm{U}(1)$ geons}

\label{sec:static}

In this section we discuss Einstein-$\mathrm{U}(1)$ geons that have a static
asymptotic region. In subsection \ref{subsec:GW-holes} we recall relevant
properties of the nondegenerate Gibbons-Wiltshire black holes with static
asymptotic regions~\cite{GibbWilt}, and in subsection \ref{subsec:GW-geons}
we construct geon versions of these holes. Subsection \ref{subsec:examples}
presents examples in the special case of constant curvature transversal
space. Generalisations beyond the Gibbons-Wiltshire spacetimes, including
generalisations to multiple (i.e., ${[\mathrm{U}(1)]}^{n}$) gauge fields,
are discussed in subsection in~\ref{subsec:generalisations}.

\subsection{Gibbons-Wiltshire metric}

\label{subsec:GW-holes}

Let $D\geq 4$, and let $(\overline{\mathcal{M}},{\overline{\mathrm{d}s^{2}}}%
) $ be a $(D-2)$-dimensional positive definite Einstein manifold: Writing $%
\overline{\mathrm{d}s^{2}}=\overline{g}_{IJ}\,\mathrm{d}y^{I}\mathrm{d}y^{J}$%
, we have $\overline{R}_{IJ}=\overline{\lambda }(D-3)\overline{g}_{IJ}$,
where $\overline{\lambda }\in \mathbb{R}$ is a constant. In local
Schwarzschild-like coordinates, the $D$-dimensional Lorentz-signature
Gibbons-Wiltshire metric reads \cite{GibbWilt}\footnote{%
For related work, see \cite%
{lemos1,lemos2,huang-liang,lemos-zanchin,mann-pair,cai-zhang,smith-mann,banados,vanzo,mann-negmass,brill-louko-peldan,birmingham-topol,emparan-johnson-myers,emparan}%
.} 
\begin{subequations}
\label{eq:GWmetric+Delta}
\begin{align}
\mathrm{d}s^{2}& =-\Delta \,\mathrm{d}t^{2}+\frac{\mathrm{d}r^{2}}{\Delta }%
+r^{2}\,{\overline{\mathrm{d}s^{2}}}\ \ ,  \label{eq:GWmetric} \\
\Delta & =\overline{\lambda }-\frac{\mu }{r^{D-3}}+\frac{Q^{2}}{r^{2(D-3)}}-%
\frac{2\Lambda r^{2}}{(D-1)(D-2)}\ \ ,
\end{align}%
where $r>0$, $\mu $ and $Q$ are real-valued constants and $\Lambda $ is the
cosmological constant. The spacetime has a Maxwell field whose Faraday
two-form is 
\end{subequations}
\begin{equation}
\boldsymbol{F}=\alpha _{D}\frac{Q}{r^{D-2}}\,\mathrm{d}t\wedge \mathrm{d}r\
\ ,  \label{eq:F-sol}
\end{equation}%
where $\alpha _{D}$ is a $D$-dependent constant whose precise value will not
be needed in what follows. Equations (\ref{eq:GWmetric+Delta}) and (\ref%
{eq:F-sol}) solve the Einstein-Maxwell equations, obtained from the action
whose gravitational part is proportional to $\int \sqrt{-g}\,(R-2\Lambda )$
and Maxwell part to $\int \sqrt{-g}\,F_{ab}F^{ab}$. When $\overline{\lambda }
$ is nonzero, it can be normalised to $\abs{\overline\lambda}=1$ without
loss of generality by rescaling $r$, $t$, $\mu $ and~$Q$.

On par with the `electric' solution given by (\ref{eq:GWmetric+Delta}) and~(%
\ref{eq:F-sol}), we consider the `magnetic' solution \cite{GibbMaeda} in
which (\ref{eq:F-sol}) is replaced by 
\begin{equation}
\boldsymbol{H}=\alpha _{D}^{\prime }Q\boldsymbol{V}_{\overline{\mathrm{d}%
s^{2}}}\ \ ,  \label{eq:H-sol}
\end{equation}%
where $\alpha _{D}^{\prime }$ is a $D$-dependent constant and $\boldsymbol{V}%
_{\overline{\mathrm{d}s^{2}}}$ is the volume form (respectively volume
density) on $(\overline{\mathcal{M}},{\overline{\mathrm{d}s^{2}}})$ if $%
\overline{\mathcal{M}}$ is orientable (nonorientable)~\cite{bott-tu}.
Equations (\ref{eq:GWmetric+Delta}) and (\ref{eq:H-sol}) 
solve the Einstein-$\mathrm{U}(1)$ equations obtained 
from the action whose gauge
field part is proportional to $\int \sqrt{-g}\,H_{a\ldots f}H^{a\ldots f}$.

Henceforth we assume
$(\overline{\mathcal{M}},{\overline{\mathrm{d}s^{2}}})$ to be
geodesically complete.

We are interested in the situation in which the metric (\ref%
{eq:GWmetric+Delta}) extends into an eternal black-and-white hole spacetime
in which a bifurcate event horizon at $\Delta =0$ \cite{wald-qft} separates
two exterior regions, both of which are static with respect to the timelike
Killing vector $\partial _{t}$ and extend to a spacelike infinity. This can
only occur for $\Lambda \leq 0$, as otherwise $\partial _{t}$ is not
timelike at large~$r$. When $\Lambda =0$, it is necessary that $\overline{%
\lambda }>0$, and the parameter range analysis and the conformal diagrams
with suppressed $(\overline{\mathcal{M}},{\overline{\mathrm{d}s^{2}}})$ are
as for Kruskal and nonextremal Reissner-Nordstr\"{o}m{}~\cite{haw-ell}. When 
$\Lambda <0$, $\overline{\lambda }$ may take any value, and the parameter
range analysis and the conformal diagrams are as in the four-dimensional
case \cite{brill-louko-peldan,lake-rnds}, with the correction pointed out in %
\cite{strobl2,Fid-etal} to those diagrams that are shown in \cite%
{brill-louko-peldan,lake-rnds} as squares.\footnote{%
We exclude the case where the spacetime would be extendible past a
coordinate singularity at $r=0$. Examination of $R_{abcd}R^{abcd}$ \cite%
{birmingham-topol} shows that this situation can only arise for $\mu =0=Q$,
and it was seen in \cite{brill-louko-peldan,birmingham-topol} that this
situation does then arise for some $(\overline{\mathcal{M}},{\overline{%
\mathrm{d}s^{2}}})$.}

Let $(\mathcal{M}, \mathrm{d} s^2)$ denote the spacetime consisting of the
four conformal blocks adjacent to the Killing horizon in the conformal
diagram. We introduce in $(\mathcal{M}, \mathrm{d} s^2)$ standard global
Kruskal null coordinates $(U,V,\boldsymbol{y})$, in which $\boldsymbol{y}$
denotes a point in~$\overline{\mathcal{M}}$. The metric reads 
\begin{equation}
\mathrm{d} s^2 = - f\, \mathrm{d} U \mathrm{d} V + r^2 \, 
{\overline{\mathrm{d} s^2}} \ \ ,  
\label{eq:kruskalmetric}
\end{equation}
the horizon is at $UV=0$, and $f$ and $r$ are smooth positive functions of
argument~$UV$. The two-form (\ref{eq:F-sol}) becomes 
\begin{equation}
\boldsymbol{F} = \alpha_D \frac{Q f}{2 \, r^{D-2}} \, \mathrm{d} U \wedge 
\mathrm{d} V \ \ ,  \label{eq:F-sol-kruskal}
\end{equation}
while the expression (\ref{eq:H-sol}) remains valid for~$\boldsymbol{H}$.
The static exteriors are at $UV<0$ and the interiors at $UV>0$. The range of 
$UV$ depends on the parameters; in particular, $UV$ is bounded below for $%
\Lambda<0$ but not for $\Lambda=0$. $(\mathcal{M}, \mathrm{d} s^2)$ may be
extendible to the future and past through further Killing horizons, but
whether or not it is, the conformal diagrams show that $(\mathcal{M}, 
\mathrm{d} s^2)$ contains all spacelike hypersurfaces that connect the two
static regions and are given in the local metric (\ref{eq:GWmetric+Delta})
by a relation between $t$ and~$r$. Working with $(\mathcal{M}, \mathrm{d}
s^2)$ will thus suffice for our purposes. If $(\mathcal{M}, \mathrm{d} s^2)$
is extendible, the quotients in the rest of this paper can be extended in an
obvious manner.

To summarise: The spacetime $(\mathcal{M}, \mathrm{d} s^2)$ is an
eternal black hole, foliated by wormhole-like spacelike hypersurfaces
of topology $\overline{\mathcal{M}} \times \mathbb{R}$. The static
regions are locally asymptotically flat for $\Lambda=0$ and locally
asymptotically anti-de~Sitter for $\Lambda<0$. 
The topology at infinity, while possibly exotic, extends to the horizon
and hence does not violate topological censorship theorems
\cite{FriedSchWi,gal-sch-witt,gal-sch-witt-wool}. 
When $(\overline{\mathcal{M}}, {\overline{\mathrm{d} s^2}})$ is the
round two-sphere and $Q=0=\Lambda$, $(\mathcal{M},
\mathrm{d} s^2)$ reduces to the Kruskal manifold.

The electric and magnetic solutions have respectively nonvanishing electric
and magnetic charges at the infinities. Suppose first that $\overline{%
\mathcal{M}}$ is orientable, in which case $(\mathcal{M},\mathrm{d}s^{2})$
is both time and space orientable and the magnetic field strength (\ref%
{eq:H-sol}) is a globally-defined $(D-2)$-form. In the magnetic solution,
the magnetic charge density on a hypersurface of constant $U$ and $V$ is the
pull-back of $\boldsymbol{H}$, and (\ref{eq:H-sol}) shows that the magnetic
charge per unit volume of $(\overline{\mathcal{M}},\overline{\mathrm{d}s^{2}}%
)$ is~$\alpha ^{\prime }Q$. In the electric solution, the electric charge
density is defined by taking the Hodge dual of $\boldsymbol{F}$ and
proceeding similarly, with the result that (with a suitable choice of
orientation) the electric charge per unit volume of $(\overline{\mathcal{M}},%
\overline{\mathrm{d}s^{2}})$ is~$\alpha Q$. {}From the globally-defined time
and space orientations of $(\mathcal{M}, \mathrm{d} s^2)$ it follows that
the charges at the two asymptopias have the opposite sign.

If $\overline{\mathcal{M}}$ is not orientable, these charge considerations
generalise by first passing to the orientable double cover of $\overline{%
\mathcal{M}}$ and then taking the quotient. For Stokes' theorem on
nonorientable manifolds in this context, see \cite%
{sorkin-relation,sorkin-multiply,fried-mayer}.

We have here presented the electric and magnetic solutions in terms of the
field strengths, rather than in terms of gauge potentials. We shall do
the same for the geon solutions in
subsection~\ref{subsec:GW-geons}.
We shall return to the gauge potentials in
section~\ref{sec:discussion}.

\subsection{Geon quotient}

\label{subsec:GW-geons}

\begin{figure}[t]
\begin{center}
\includegraphics[width=0.5\textwidth]{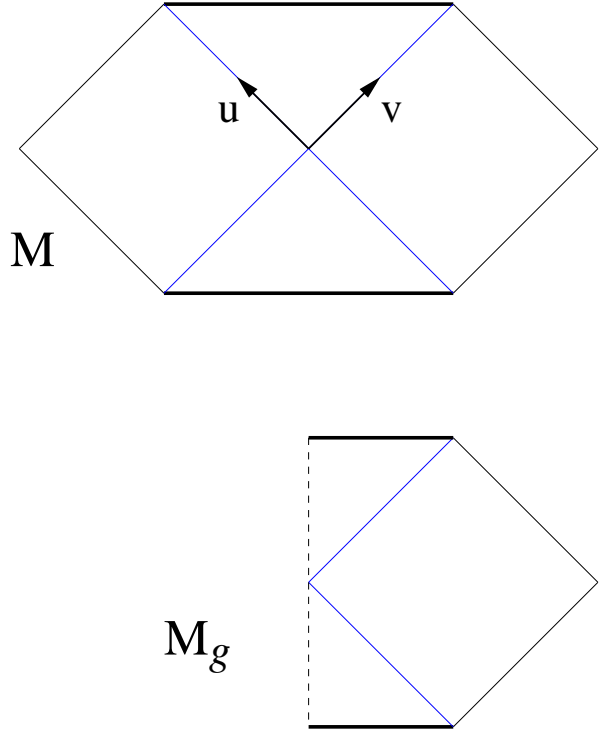}
\end{center}
\caption{A conformal diagram of the case in which 
the static regions of 
$\mathcal{M}$ are asymptotically locally flat and the nonstatic
regions terminate at spacelike singularities. 
The Kruskal null coordinates in (\ref{eq:kruskalmetric}) are
chosen to have the range $-\infty < UV < 1$, with the infinities at
$UV \to -\infty$ and the singularities at $UV \to 1$, and the null
coordinates $(u,v)$ of the conformal diagram are obtained by $U =
\tan(u)$ and $V = \tan(v)$. The conformal diagram of the quotient
spacetime $\mathcal{M}_{g}:=
\mathcal{M}/\Gamma$ 
is the (say) right half of that of 
$\mathcal{M}$, with generic points in the diagram
representing a suppressed 
$\overline{\mathcal{M}}$
but the points on the dashed line on the left representing a
suppressed $\overline{\mathcal{M}}/\overline{\Gamma}$. 
}
\label{fig:conformal}
\end{figure}

Suppose that $(\overline{\mathcal{M}},{\overline{\mathrm{d}s^{2}}})$ admits
a freely-acting involutive isometry~$P$, and let 
$\overline{\Gamma}:=
\bigl\{\mathrm{Id}_{\overline{\mathcal{M}}},P\bigr\}
\simeq \mathbb{Z}_{2}$ be the isometry group generated by~$P$.
The map
\begin{equation}
J:\bigl(U,V,\boldsymbol{y}\bigr)\mapsto \bigl(V,U,P(\boldsymbol{y})\bigr)
\label{eq:J-action}
\end{equation}
is then a freely-acting involutive isometry on
$(\mathcal{M},\mathrm{d}s^{2})$, and it generates the isometry group
$\Gamma :=\bigl\{\mathrm{Id}_{\mathcal{M}},J\bigr\}\simeq
\mathbb{Z}_{2}$. $J$~acts in the conformal diagram by the 
left-right reflection in the displayed two dimensions and by $P$ in the
suppressed $D-2$ dimensions, from which it is seen that $J$ preserves
the time orientation on $(\mathcal{M},\mathrm{d}s^{2})$ and maps the
two static regions to each other. The case where the static regions
are asymptotically locally 
flat and the nonstatic regions end at a spacelike 
singularity is
shown in Figure~\ref{fig:conformal}.

The quotient $\bigl(\mathcal{M}_{g},\mathrm{d}s_{g}^{2}\bigr):=
(\mathcal{M}, \mathrm{d}s^{2})/\Gamma$ is a geon. For $Q=0$,
$\bigl(\mathcal{M}_{g}, \mathrm{d}s_{g}^{2}\bigr)$ clearly satisfies
Einstein's equations. For $Q\neq 0$, there is a sublety in that the
two-form (\ref{eq:F-sol-kruskal}) changes sign under~$J$, and the
$(D-2)$-density (\ref{eq:H-sol}) may change sign under $J$ depending
on the $P$-action on
$(\overline{\mathcal{M}},{\overline{\mathrm{d}s^{2}}})$. Hence
$\bigl(\mathcal{M}_{g},\mathrm{d}s_{g}^{2}\bigr)$ with the quotient
gauge field solves the Einstein-$\mathrm{U}(1)$ field equations iff
the field strength can be interpreted as a section of a bundle 
whose twisting is compatible with these sign changes.
We now show that the field strength
has this interpretation and describe the
bundles in terms of $\pi _{1}\bigl(\mathcal{M}_{g}\bigr)$.

Let $(\widetilde{\mathcal{M}}, {\widetilde{\mathrm{d} s^2}})$ be the
universal covering space of $(\mathcal{M}, \mathrm{d} s^2)$, and let $\bigl(%
\widetilde{\overline{\mathcal{M}}}, {\widetilde{\overline{\mathrm{d} s^2}}}%
\bigr)$ be the universal covering space of $\bigl(\overline{\mathcal{M}}, {%
\overline{\mathrm{d} s^2}}\bigr)$. On $\bigl(\widetilde{\mathcal{M}}, {%
\widetilde{\mathrm{d} s^2}}\bigr)$, the metric is obtained from (\ref%
{eq:kruskalmetric}) by the replacements $\mathrm{d} s^2 \to {\widetilde{%
\mathrm{d} s^2}}$ and ${\overline{\mathrm{d} s^2}} \to {\widetilde{\overline{%
\mathrm{d} s^2}}}$, $\boldsymbol{F}$ is given by~(\ref{eq:F-sol}), and $%
\boldsymbol{H}$ by 
\begin{equation}
\boldsymbol{H} = \alpha^{\prime}_D Q \boldsymbol{V}_{\widetilde{\overline{%
\mathrm{d} s^2}}} \ \ .  \label{eq:H-sol-cover}
\end{equation}
Since $\widetilde{\overline{\mathcal{M}}}$ is orientable, $\boldsymbol{H}$ (%
\ref{eq:H-sol-cover}) is a globally-defined $(D-2)$-form on~$\widetilde{%
\mathcal{M}}$.


It follows from the properties of fundamental groups of quotient spaces \cite%
{hatcher} that $\bigl(\mathcal{M}_g, \mathrm{d} s^2_g\bigr) = \bigl(%
\widetilde{\mathcal{M}}, {\widetilde{\mathrm{d} s^2}}\bigr)/G$ and $(%
\mathcal{M}, \mathrm{d} s^2) = \bigl(\widetilde{\mathcal{M}}, {\widetilde{%
\mathrm{d} s^2}}\bigr)/H$, where $G\simeq \pi_1\bigl(\mathcal{M}_g\bigr)$ is
a discrete group of isometries of $\bigl(\widetilde{\mathcal{M}}, {%
\widetilde{\mathrm{d} s^2}}\bigr)$, acting freely and properly
discontinuously, $H\simeq \pi_1(\mathcal{M}) \simeq \pi_1 \bigl(\overline{%
\mathcal{M}}\bigr)$ is a normal subgroup of $G$ and $G/H \simeq \Gamma
\simeq \mathbb{Z}_2$. By restricting the action of $G$ to the invariant
submanifold at $U=0=V$, we obtain from the groups $H \subsetneq G$ the
isomorphic groups $\overline{H} \subsetneq \overline{G}$ of isometries of $%
\bigl(\widetilde{\overline{\mathcal{M}}}, {\widetilde{\overline{\mathrm{d}
s^2}}}\bigr)$. It follows that the action of $\overline{G}$ is free and
properly discontinuous, $\bigl(\overline{\mathcal{M}}, \overline{\mathrm{d}
s^2}\bigr)/\overline{\Gamma} = \bigl(\widetilde{\overline{\mathcal{M}}}, {%
\widetilde{\overline{\mathrm{d} s^2}}}\bigr)/\overline{G}$ and $\bigl(%
\overline{\mathcal{M}}, \overline{\mathrm{d} s^2}\bigr) = \bigl(\widetilde{%
\overline{\mathcal{M}}}, {\widetilde{\overline{\mathrm{d} s^2}}}\bigr)/%
\overline{H}$.

Let $\overline{G}_{+}\subset \overline{G}$ be the normal subgroup that
preserves the orientation on~$\widetilde{\overline{\mathcal{M}}}$, and let $%
G_{+}\subset G$ be the corresponding subgroup defined by the isomorphism $%
\overline{G}\simeq G$. $\overline{G}/\overline{G}_{+}$ is then trivial if $%
\overline{\mathcal{M}}/\overline{\Gamma }$ is orientable and isomorphic to $%
\mathbb{Z}_{2}$ if $\overline{\mathcal{M}}/\overline{\Gamma }$ is
nonorientable. The following proposition shows that three qualitatively
different situations arise.

\begin{proposition}
\label{prop:subgroups} The subgroups $\overline{H}\subsetneq \overline{G}$
and $\overline{G}_{+}\subset \overline{G}$ can occur in the following three
combinations:

\begin{itemize}
\item[(i)] $\overline{\mathcal{M}}$ is orientable and $P$ reverses its
orientation. Then $\overline{H} = \overline{G}_+ \subsetneq \overline{G}$. $%
\mathcal{M}_g$~is orientable.

\item[(ii)] $\overline{\mathcal{M}}$ is orientable and $P$ preserves its
orientation. Then $\overline{H} \subsetneq \overline{G}_+ = \overline{G}$. $%
\mathcal{M}_g$~is nonorientable.

\item[(iii)] $\overline{\mathcal{M}}$ is nonorientable. Then $\overline{H}%
\cap \overline{G}_{+}\subset \overline{G}$ is a normal subgroup and $%
\overline{G}/(\overline{H}\cap \overline{G}_{+})$ is canonically isomorphic
to $(\overline{G}/\overline{H})\times (\overline{G}/\overline{G}_{+})\simeq 
\mathbb{Z}_{2}\times \mathbb{Z}_{2}$. $\mathcal{M}_{g}$~is nonorientable.
\end{itemize}
\end{proposition}

\emph{Proof\/}. Types (i) and (ii) are clear. In Type (iii), the subgroups $%
\overline{H} \subsetneq \overline{G}$ and $\overline{G}_+ \subsetneq 
\overline{G}$ partition $\overline{G}$ into four subsets, at least three of
which are nonempty by assumption. Examination of the quotient groups $%
\overline{G}/\overline{H} \simeq \mathbb{Z}_2 \simeq \overline{G}/\overline{G%
}_+$ shows that all four are nonempty and establishes the isomorphism. $%
\blacksquare$

Consider now the gauge field configurations in view of this structure. In
the quotient of 
$\bigl(\widetilde{\mathcal{M}},{\widetilde{\mathrm{d}s^{2}}}\bigr)$ by~$G$, 
the electric 
field strength 
(\ref{eq:F-sol}) needs to be
twisted by~$G/H$, while the magnetic 
field strength 
(\ref{eq:H-sol-cover})
needs to be twisted by~$G/G_{+}$, where in each case the nontrivial 
$\mathbb{Z}_{2}$ element multiplies the form by~$-1$. 
The isomorphism $G\simeq \pi_{1}\bigl(\mathcal{M}_{g}\bigr)$ 
shows that this twisting defines the 
field strength as a 
section of a bundle over~$\mathcal{M}_{g}$. 
Note that as $\bigl(\mathcal{M}_{g},\mathrm{d}s_{g}^{2}\bigr)$ 
has a global foliation by
spacelike hypersurfaces, 
$\pi_{1}\bigl(\mathcal{M}_{g}\bigr)$ is isomorphic to the 
spatial fundamental group of~$\mathcal{M}_{g}$. 

All the electric and magnetic geons have respectively a nonvanishing
electric and magnetic charge at the single asymptopia. This is not forbidden
by Stokes' theorem because of the nontriviality of the 
field strength bundle and/or the role
of (non)orientability in Stokes' theorem \cite%
{sorkin-relation,sorkin-multiply,fried-mayer}: the only untwisted
field strength occurs for a Type (ii) magnetic geon, which is nonorientable.%
\footnote{%
It was noted in \cite{sorkin-relation,sorkin-multiply,fried-mayer} that an
untwisted magnetic field strength on a nonorientable manifold with one
asymptopia can have nonvanishing charge. That the $\mathbb{RP}^{3}$ geon
does not generalise to accommodate an untwisted electric field was noted in
terms of the initial data already in~\cite{MW-geondata}.} 
Note that the 
only field
strength that is twisted by the (spatial) orientation bundle of 
$\mathcal{M}_{g}$ occurs in a Type (ii) electric geon.

In the special case $D=4$, in which the electric and magnetic 
field strengths
are both locally two-forms, we can obtain further geons by starting from a
linear combination. On the one hand, such constructions are constrained by
the requirement that the electric and magnetic parts 
of the field strength must be sections of the
same bundle (so that the 
linear combination is well-defined). But, on the
other hand, we have the opportunity obtain further geons by using the Hodge
dual to twist the 
bundle. Let $\boldsymbol{F}_{0}$ and $\boldsymbol{H}_{0}$
stand for respectively (\ref{eq:F-sol}) and (\ref{eq:H-sol-cover}) on $\bigl(%
\widetilde{\mathcal{M}},{\widetilde{\mathrm{d}s^{2}}}\bigr)$. The linear
combination 
\begin{equation}
\boldsymbol{F}_{\beta }:=\cos (\beta )\boldsymbol{F}_{0}+\sin (\beta )%
\boldsymbol{H}_{0}  \label{eq:Fbeta-sol-cover}
\end{equation}%
solves then the two-form Einstein-$\mathrm{U}(1)$ equations on $\bigl(%
\widetilde{\mathcal{M}},{\widetilde{\mathrm{d}s^{2}}}\bigr)$ for any $\beta
\in \mathbb{R}$, and we may understand $\beta $ identified with period $2\pi 
$. Fix now the orientation on~$\widetilde{\mathcal{M}}$, for concreteness
but without loss of generality so that $\ast \boldsymbol{F}_{\beta }=%
\boldsymbol{F}_{\beta +\pi /2}$. For the three types in Proposition~\ref%
{prop:subgroups}, we then obtain solutions as follows:

\textit{Type~(i)\/}. Since $H=G_{+}$, no additional constraints arise from
taking the electric and magnetic parts 
of the field strength to be sections of the same bundle.
Thus, $\beta$ is arbitrary and the nontrivial element of $G/H=G/G_{+}\simeq 
\mathbb{Z}_{2}$ multiplies the field strength
by~$-1$. However, the Hodge dual
offers no further inequivalent possibilities.

\textit{Type~(ii)\/}. $\beta=\pi/4$ or $\beta=5\pi/4$ and the nontrivial
element of $G/H\simeq\mathbb{Z}_2$ acts on the field strength
by the Hodge dual.  
There is no contradiction with the fact that the Hodge dual
squares to minus identity, since the nontrivial element of $G/H$
reverses the orientation on $\widetilde{\mathcal{M}}$ and the Hodge
dual acquires an additional minus sign on orientation reversal.

\textit{Type~(iii)\/}. $\beta=\pi/4$ or $\beta=5\pi/4$, the nontrivial
element of $G/H$ acts on the field strength
by the Hodge dual and the nontrivial
element of $G/G_+$ by the inverse of the Hodge dual. As the nontrivial
elements of both $G/H$ and $G/G_+$ reverse the orientation 
on~$\widetilde{\mathcal{M}}$, 
there is again no contradiction with the Hodge dual squaring
to minus identity.

\subsection{Examples: Constant curvature transversal space}

\label{subsec:examples}

We have seen that classifying geons within the Gibbons-Wiltshire metrics
reduces to classifying quotients of the transversal space $\bigl(\widetilde{%
\overline{\mathcal{M}}},{\widetilde{\overline{\mathrm{d}s^{2}}}}\bigr)$. We
now examine these quotients in the special case of constant curvature $\bigl(%
\widetilde{\overline{\mathcal{M}}},\widetilde{\overline{\mathrm{d}s^{2}}}%
\bigr)$.

\subsubsection*{$\overline{\protect\lambda}=1$: Spherical spaces}


When $\overline{\lambda }=1$,
$\bigl(\widetilde{\overline{\mathcal{M}}},
{\widetilde{\overline{\mathrm{d}s^{2}}}}\bigr)$ 
is $(D-2)$-dimensional round
sphere, $S^{D-2}$. The quotients are classified in~\cite{wolf}. For
any~$D$, we can take $\overline{G}$ to be the order two group
generated by the antipodal map: This is within our Type (i) for even
$D$ and Type (ii) for odd~$D$. For even $D$ there are no other
possibilities. For odd~$D$, $S^{D-2} $ has infinitely many other
quotients satisfying our subgroup assumptions, all of them
Type~(ii).\footnote{%
Type (ii) is the only possibility because all
quotients of odd-dimensional spheres are orientable. A~reader who
wishes to forgo the pleasure of verifying this from the full list of
quotients in \cite{wolf} may prefer the following argument. Let $D$ be
odd and let $S^{D-2}/L$ be a spherical space, where $L\subset
\mathrm{O}(D-1)$ is a subgroup with a free and properly discontinuous
action. Suppose $l\in L$ is an element that reverses the orientation
of $S^{D-2}$, and let $L_{0}\subset L$ be the subgroup generated
by~$l$. Then $S^{D-2}/L_{0}$ is a spherical space. If $L_{0}$ has
order~$2$, it follows from page 218 in \cite{wolf} that $l$ is the
antipodal map, but this is a contradiction since the antipodal map for
odd $D$ preserves the orientation. If $L_{0}$ has order greater
than~$2$, it falls into type I of~\cite{wolf}. In this case equation
(7.4.1) and Theorems 5.5.6 and 5.5.11 in
\cite{wolf} show that all elements in $L_{0}$ preserve the orientation,
which is again a contradiction.} 
The simplest examples arise 
when 
$\overline{G}$ is a cyclic group of even order and 
$\overline{H}$ is the subgroup that
consists of even powers of the generator; 
other examples arise by setting the angular momenta
in the rotating geons of subsection \ref{subsec:J-on-spheres} to
zero. Note that the only case in which the spatial
topology at infinity is the conventional $S^{D-3}$ is when $\overline{G}$ is
generated by the antipodal map.

When $D=4$ and the form fields vanish, the only possibility is the $\mathbb{%
RP}^{3}$ geon \cite{MW-geondata,FriedSchWi,giulini-thesis,giulini-multiRP3s}
for $\Lambda =0$ and its asymptotically anti-de~Sitter generalisation for $%
\Lambda <0$.

\subsubsection*{$\overline{\protect\lambda}=0$: Flat spaces}


When $\overline{\lambda }=0$, $\bigl(\widetilde{\overline{\mathcal{M}}},{%
\widetilde{\overline{\mathrm{d}s^{2}}}}\bigr)$ is flat $\mathbb{R}^{D-2}$.
Quotients of Types (i), (ii) and (iii) exist for all~$D$. For $D=4$, compact
examples are obtained when the map $\bigl(\overline{\mathcal{M}},{\overline{%
\mathrm{d}s^{2}}}\bigr)\rightarrow \bigl(\overline{\mathcal{M}},{\overline{%
\mathrm{d}s^{2}}}\bigr)/\overline{\Gamma }$ is respectively (i) double
covering of a Klein bottle by a torus, (ii) double covering of a torus by a
torus, and (iii) double covering of a Klein bottle by a Klein bottle. Taking
a product with a flat $(D-3)$-torus produces compact examples for
$D>4$.
Other examples arise by setting the angular momenta in the
rotating geons of subsection \ref{subsec:J-on-tori} to zero.

\subsubsection*{$\overline{\protect\lambda}=-1$: Hyperbolic spaces}


When $\overline{\lambda }=-1$, $\bigl(\widetilde{\overline{\mathcal{M}}},{%
\widetilde{\overline{\mathrm{d}s^{2}}}}\bigr)$ is $(D-2)$-dimensional
hyperbolic space. Quotients of Types (i), (ii) and (iii) exist for all~$D$.
For $D>4$, noncompact examples arise by writing the metric in Poincare
coordinates, 
\begin{equation}
{\widetilde{\overline{\mathrm{d}s^{2}}}}=\frac{\mathrm{d}z^{2}+\mathrm{d}%
\mathbf{x}^{2}}{z^{2}}\ \ ,
\end{equation}%
where $z>0$ and $\mathrm{d}\mathbf{x}^{2}$ is the standard flat metric on $%
\mathbb{R}^{D-3}$, and letting $\overline{G}$ act on $\mathbf{x}\in \mathbb{R%
}^{D-3}$ as in the $\overline{\lambda }=0$ case. For $D=4$, an example of
Type (ii) arises when $\overline{G}$ is generated by a hyperbolic or
elliptic M\"{o}bius transformation, and an example of Type (i) arises when $%
\overline{G}$ is generated by a glide-reflection, a hyperbolic M\"{o}bius
transformation followed by the reflection about the axis of this
transformation. An example of Type (iii) arises when $\overline{G}$ is
generated by a glide-reflection $A$ and a hyperbolic or elliptic M\"{o}bius
transformation~$B$, with the parameters chosen so that the group acts freely
(cf.\ the pairs of hyperbolic elements analysed in~\cite{horo-maro}), and $%
\overline{H}$ is generated by $A$ and~$B^{2}$. Compact $D=4$ examples are
discussed in \cite{schiffer-spencer,alling-greenleaf}.

\subsection{Generalisations}

\label{subsec:generalisations}

We now consider some generalisations beyond the Gibbons-Wiltshire spacetimes.

The quotient technique generalises immediately to brane-like solutions in
which the brane dimensions are sufficiently inert. An example in vacuum with 
$\Lambda =0$ is the product of Kruskal with any positive definite Ricci-flat
space. The brane dimensions may offer new choices for the isometry~$P$, and
this freedom may in particular be used to make spatially orientable or
nonorientable geons as desired. As an example, in the stringy black hole
that is the product of the spinless BTZ hole and $S^{3}\times T^{4}$,
choosing $P$ to have a suitable nontrivial action on the $T^{4}$ factor
gives a spatially orientable geon~\cite{louko-marolf}.

The technique also generalises immediately to form fields of more general
rank with a pure $\boldsymbol{F}^{2}$ action. A~generalisation to ${[\mathrm{%
U}(1)]}^{n}$ form fields is possible and gives new opportunities to twist
the bundle by permuting the $\mathrm{U}(1)$ components. Terms other than $%
\boldsymbol{F}^{2}$ in a form field action may however bring about new
phenomena. For example, a Chern-Simons term, such as that in 11-dimensional
supergravity~\cite{duff-nilsson-pope}, must be a top rank form when the
spacetime is orientable but a top rank density when the spacetime is
nonorientable. Finally, including dilatonic scalar fields, such as those in %
\cite{GibbMaeda,ChanHorneMann}, is straightforward.

There exist generalisations without an asymptotic region of the kind we have
assumed above. One example is the Bertotti-Robinson-type extremal limit of~(%
\ref{eq:GWmetric+Delta}), in which the spacetime is the product of $(1+1)$%
-dimensional anti-de~Sitter space and a constant multiple of $(\overline{%
\mathcal{M}},{\overline{\mathrm{d}s^{2}}})$. The horizon is then a Killing
horizon on the anti-de~Sitter space, and the quotient analysis proceeds as
in subsection~\ref{subsec:GW-geons}. In the limit of vanishing curvature,
these spacetimes reduce to geon-like versions of Rindler space~\cite%
{louko-marolf-rp3}.

Another example is de~Sitter space. The geon-like quotient yields a version
in which the spacelike hypersurfaces in the global foliation are not $%
S^{D-1} $ but $S^{D-1}/\mathbb{Z}_2 \simeq \mathbb{RP}^{D-1}$ \cite%
{louko-schleich,schleich-witt-rp3,McInnes-schwds}.

Yet another example occurs when $\Lambda >0$ and the conformal diagram is as
for the nondegenerate Schwarzschild-de~Sitter solution, with a pattern of
black hole and cosmological horizons repeating infinitely in the horizontal
direction~\cite{gh-deS}. It is possible to take a $\mathbb{Z}_{2}$ quotient
using one of the horizons, either a black hole horizon or a cosmological
one. The conformal diagram becomes then bounded from (say) the left and
infinite to the right. However, it is also possible to take a quotient under
a group generated by two involutive isometries, each using a different
horizon, so that the conformal diagram becomes bounded both from the left
and from the right~\cite{McInnes-schwds}. In this case the fixed timelike
hypersurfaces of the two isometries can be arbitrarily boosted with respect
to each other (cf.\ the discussion in two-dimensional dilaton gravity
context in~\cite{strobl3}, Figure~14). These quotients might offer an arena
for probing dS/CFT correspondence \cite{witten-dscft,strominger-dscft}.

Finally, the higher-dimensional BTZ hole \cite%
{banados,aminneborg,holst,CreightonMann,gomberoff} resembles a geon in that
the exterior region is connected. It is however possible to take a $\mathbb{Z%
}_{2}$ quotient of this hole, with the effect that the topology at infinity
changes and the isometry group becomes smaller. These properties were
investigated in the AdS/CFT context in~\cite{louko-wisniewski}.

\section{Angular momentum}

\label{sec:angmom}

In this section we construct rotating geons. We consider first angular
momenta on a torus and then angular momenta on a (non-round) sphere.

\subsection{Torus}

\label{subsec:J-on-tori}

Let $(\mathcal{N},\mathrm{d}s^{2})$ be the planar vacuum Gibbons-Wiltshire
hole with $\Lambda <0$. The metric reads 
\begin{equation}
\mathrm{d}s^{2}=-f\,\mathrm{d}U\mathrm{d}V+r^{2}\,\mathrm{d}\ell ^{2}\ \ ,
\end{equation}%
where $\mathrm{d}\ell ^{2}$ is the standard flat positive definite metric on 
$\mathbb{R}^{D-2}$ and $f$ and the range of the Kruskal coordinates $(U,V)$
are as in (\ref{eq:kruskalmetric}) with $\Lambda <0=\overline{\lambda }$.
Let $1\leq d\leq D-2$, let $\bigl\{\boldsymbol{\xi}^{(\alpha )}\mid \alpha
=1,\ldots ,d\bigr\}$ be a linearly independent set of translational Killing
vectors on $(\mathbb{R}^{D-2},\mathrm{d}\ell ^{2})$, and let 
\begin{equation}
\boldsymbol{\eta}^{(\alpha )}:=\boldsymbol{\xi}^{(\alpha )}+a^{(\alpha
)}(V\partial _{V}-U\partial _{U})\ \ ,
\end{equation}%
where $a^{(\alpha )}$ are real-valued constants, assumed so small in
absolute value that $\boldsymbol{\eta}^{(\alpha )}\cdot \boldsymbol{\eta}%
^{(\alpha )}\rightarrow \infty $ as $r\rightarrow \infty $ for each~$\alpha $%
. $\bigl\{\boldsymbol{\eta}^{(\alpha )}\bigr\}$ is a set of commuting
Killing vectors on $(\mathcal{N},\mathrm{d}s^{2})$. The group $H_{0}$
generated by the exponential maps of these Killing vectors acts freely and
properly discontinuously, and the quotient $(\mathcal{N},\mathrm{d}%
s^{2})/H_{0}=:(\mathcal{N}_{0},\mathrm{d}s_{0}^{2})$ is an eternal black
hole with spatial topology $T^{d}\times \mathbb{R}^{D-2-d}$.

When all the $a$'s vanish, we are in the situation covered in 
section~\ref{sec:static}. 
Assume from now on that at least one of the $a$'s is
nonvanishing. $(\mathcal{N}_0, \mathrm{d} s^2_0)$ 
is then a spinning eternal
black hole and the $a$'s 
have an interpretation in terms of angular momenta
\cite{lemos2,lemos-zanchin,awad}.

In taking a geon quotient of $(\mathcal{N}_0, \mathrm{d} s^2_0)$ we face a
new difficulty in that the map $(U,V)\mapsto (V,U)$ on 
$(\mathcal{N},\mathrm{d}s^{2})$ reverses the signs of the~$a$'s. 
The angular momenta at the two
infinities, defined with respect to the global time orientation on 
$(\mathcal{N}_{0},\mathrm{d}s_{0}^{2})$, 
therefore have opposite signs. There
do exist time-orientation preserving isometries between the two exterior
regions, but in general such an isometry needs to invert all the spatial
dimensions, which implies that the isometry does not extend into an
involutive fixed-point free isometry of $(\mathcal{N}_{0},\mathrm{d}%
s_{0}^{2})$.

However, in certain special configurations a geon quotient exists. We give
two examples.

First, suppose that $d\ge2$, $\boldsymbol{\xi}^{(1)}$ is orthogonal to all
others $\boldsymbol{\xi}$'s and $a^{(1)}=0$. We introduce on $(\mathbb{R}%
^{D-2}, \mathrm{d} \ell^2)$ the adapted coordinates $\boldsymbol{y} = (y^1, %
\boldsymbol{y}_\perp) \in \mathbb{R} \times \mathbb{R}^{D-3} \simeq \mathbb{R%
}^{D-2}$, in which $\boldsymbol{\xi}^{(1)} = b \partial_{y^1}$. Consider on $%
(\mathcal{N}, \mathrm{d} s^2)$ the isometry 
\begin{equation}
J_1: \bigl( U,V, y^1, \boldsymbol{y}_\perp \bigr) \mapsto \bigl(V,U, y^1 +
(b/2), -\boldsymbol{y}_\perp \bigr) \ \ ,
\end{equation}
which preserves $\boldsymbol{\eta}^{(1)}$ and reverses the signs of all
other~$\boldsymbol{\eta}$'s. Let $G_0$ be the group generated by $H_0$ and~$%
J_1$. $G_0$ acts on $(\mathcal{N}, \mathrm{d} s^2)$ freely and properly
discontinuosly, $H_0 \subset G_0$ is a normal subgroup with $G_0/H_0 \simeq 
\mathbb{Z}_2$ and $(\mathcal{N}, \mathrm{d} s^2)/G_0 \simeq (\mathcal{N}_0, 
\mathrm{d} s^2_0)/\mathbb{Z}_2$ is a geon. There can be up to $D-3$
independent angular momentum parameters.

Second, suppose $D\geq 5$, $d\geq 3$, $\boldsymbol{\xi}^{(1)}$ and $%
\boldsymbol{\xi}^{(2)}$ have equal magnitude and are orthogonal to all other 
$\boldsymbol{\xi}$'s and the only nonvanishing $a$'s are 
$a^{(1)}=-a^{(2)}$.
Introduce on $(\mathbb{R}^{D-2},\mathrm{d}\ell ^{2})$ the adapted
coordinates $\boldsymbol{y}=\bigl((y^{1},y^{2}),\boldsymbol{y}_{\perp }\bigr)%
\in \mathbb{R}^{2}\times \mathbb{R}^{D-4}\simeq \mathbb{R}^{D-2}$, 
in which 
$\boldsymbol{\xi}^{(1)}=b\partial _{y^{1}}$ 
and 
$\boldsymbol{\xi}^{(2)}=b\partial _{y^{2}}$. 
(Note that $\boldsymbol{\xi}^{(1)}$ and 
$\boldsymbol{\xi}^{(2)}$ need not be orthogonal.) 
Consider on $(\mathcal{N},\mathrm{d}s^{2})$ the isometry 
\begin{equation}
J_{2}:\bigl(U,V,y^{1},y^{2},\boldsymbol{y}_{\perp }\bigr)\mapsto \bigl(%
V,U,y^{2},y^{1},\exp (\tfrac{1}{2}\boldsymbol{\xi}^{(3)})\boldsymbol{y}%
_{\perp }\bigr)\ \ ,  \label{eq:J2}
\end{equation}%
which interchanges $\boldsymbol{\eta}^{(1)}$ and $\boldsymbol{\eta}^{(2)}$
and preserves all other~$\boldsymbol{\eta}$'s. The quotient of $(\mathcal{N}%
, \mathrm{d} s^2)$ by the group generated by $H_{0}$ and $J_{2}$ is then a $%
\mathbb{Z}_{2}$ geon quotient of $(\mathcal{N}_{0},\mathrm{d}s_{0}^{2})$.

For some block-diagonal configurations there exist $\mathbb{Z}_2$ quotients
in which the involutive isometry acts as with $J_1$ in some blocks and as
with $J_2$ in others. Also, for some configurations further rotating geons
may be constructed by a quotient that has order higher than $2$, using maps
that generalise $J_2$ to permute an even number of dimensions. We shall not
attempt to enumerate all the possibilities here.

Dilatonic fields and antisymmetric tensor fields can be included as with the
nonrotating geons.

\subsection{Sphere}

\label{subsec:J-on-spheres}

We consider in $D\ge4$ dimensions the rotating Myers-Perry(-AdS) 
black holes 
\cite{myers-perry,hawking-hunter-tr,gibbons-lu-page-pope1,%
gibbons-lu-page-pope2,gibbons-perry-pope},
which solve the vacuum Einstein equations with a vanishing
(respectively negative) cosmological constant. We assume positive mass
and a nondegenerate horizon. We use the Kruskal coordinates that were
explicitly given for a vanishing cosmological constant in
\cite{myers-perry} and whose construction for a negative cosmological
constant is similar.

In the Kruskal coordinates of~\cite{myers-perry}, a left-right reflection in
the conformal diagram reverses the signs of all the angular momenta. An
isometry $K$ that involves such a reflection thus needs to include some
compensating operation in the $D-2$ dimensions suppressed in the diagram. In
order that $K$ can be used in the quotient, the group generated by $K$ needs
to act freely and properly discontinuously. If the quotient is further
required to be continuously deformable to the nonrotating case, no such $K$
exists for even $D$ or for $D=7$. For even~$D$, the reason is that in the
nonrotating limit $\overline{G}$ must be the order two group generated by
the antipodal map, as discussed in subsection~\ref{subsec:examples}: as the
nontrivial element of $G$ then leaves all the (nonrotating) rotation planes
invariant, this element cannot be the limit of an isometry that permutes
(rotating) rotation planes. For $D=7$, the reason is that in the nonrotating
limit $\overline{G}$ must be either the order two group generated by the
antipodal map or one of the higher order isometry groups of the round
5-sphere listed on p.~225 in~\cite{wolf}. The only nontrivial action of $G$
on the three commuting (nonrotating) rotation planes is then a cyclic
permutation, and this cannot be the limit of an action on a rotating
spacetime since a cyclic permutation of an odd number of angular momenta is
equivalent to an overall sign change only when all the angular momenta in
fact vanish.

When $D$ is odd and not equal to~$7$, rotating geons exist. We give an
example.

First, suppose $D=4p+1$ where $p$ is a positive integer. We define the null
Kruskal coordinates $(U,V)$ by $U:=v-u$ and $V:=v+u$, where $u$ and $v$ are
respectively the spacelike and timelike Kruskal coordinates of~\cite%
{myers-perry}. We denote the first $p$ rotation angles by $%
\boldsymbol{\theta}=(\theta _{1},\ldots ,\theta _{p})$, the associated
direction cosines by $\boldsymbol{\mu}=(\mu _{1},\ldots ,\mu _{p})$, the
remaining $p$ rotation angles by $\boldsymbol{\varphi}=(\varphi _{1},\ldots
,\varphi _{p})$ and the associated direction cosines by $\boldsymbol{\nu}%
=(\nu _{1},\ldots ,\nu _{p})$. Recall that all the direction cosines are
non-negative and satisfy $\boldsymbol{\mu}^{2}+\boldsymbol{\nu}^{2}=1$, each
rotation angle is periodic with period $2\pi $, and that coordinate
singularities only occur when the vanishing of a direction cosine makes the
corresponding rotation angle ambiguous.

Assume now the angular momentum in each $\theta _{i}$ to be the negative of
the angular momentum in the corresponding~$\varphi _{i}$. Let $q$ and $r_{i}$%
, $1\leq i\leq p$, be positive integers such that $\gcd (r_{i},2q)=1$ for
all~$i$. Consider the map 
\begin{equation}
K_{1}:\bigl(U,V,\boldsymbol{\mu},\boldsymbol{\nu},\boldsymbol{\theta},%
\boldsymbol{\varphi}\bigr)\mapsto \bigl(V,U,\boldsymbol{\nu},\boldsymbol{\mu}%
,\boldsymbol{\varphi},\boldsymbol{\theta}+(\pi /q)\boldsymbol{r}\bigr)\ \ ,
\end{equation}%
where $\boldsymbol{r}:=(r_{1},\ldots ,r_{p})$. \ In words, $K_{1}$ effects a
left-right reflection in the conformal diagram, interchanges all the paired
rotation planes, and rotates in the last $p$ rotation planes by~$\pi r_{i}/q$%
. $K_{1}$~generates an isometry group of order~$4q$, this group acts freely
and properly discontinuously, and the quotient is a geon. The geon is
asymptotically locally flat, the spatial infinity being an order $2q$
quotient of $S^{4p-1}$ and reducing to $\mathbb{RP}^{4p-1}$ for~$q=1$. The
geon can be written as the $\mathbb{Z}_{2}$ quotient of the two-exterior
hole that is the quotient of the original spacetime under the order $q$
isometry group generated by~$K_{1}^{2}$.

Second, suppose $D=4p+7$ where $p$ is a positive integer. We denote the
first $2p$ rotation angles and direction cosines as above, the remaining
three rotation angles by $(\psi _{1},\psi _{2},\psi _{3})$ and their
direction cosines by $(\rho _{1},\rho _{2},\rho _{3})$. The direction
cosines are non-negative and satisfy $\boldsymbol{\mu}^{2}+\boldsymbol{\nu}%
^{2}+\rho _{1}^{2}+\rho _{2}^{2}+\rho _{3}^{2}=1$, and the periodicities and
coordinate singularities are as above. Assume again the angular momentum in
each $\theta _{i}$ to be the negative of the angular momentum in the
corresponding~$\varphi _{i}$, and assume in addition the angular momentum in
each $\psi _{i}$ to vanish. Let $q$ and $s$ be positive integers satisfying $%
4q=9s$, let $r_{i}$ be as above, and let $t$ be a positive integer
satisfying $\gcd (t,3s)=1$. Consider the map 
\begin{align}
& K_{2}:\bigl(U,V,\boldsymbol{\mu},\boldsymbol{\nu},\rho _{1},\rho _{2},\rho
_{3},\boldsymbol{\theta},\boldsymbol{\varphi},\psi _{1},\psi _{2},\psi _{3}%
\bigr)\mapsto  \notag \\
& \hspace{7ex}\bigl(V,U,\boldsymbol{\nu},\boldsymbol{\mu},\rho _{2},\rho
_{3},\rho _{1},\boldsymbol{\varphi},\boldsymbol{\theta}+(\pi /q)%
\boldsymbol{r},\psi _{2},\psi _{3},\psi _{1}+2\pi t/(3s)\bigr)\ \ .
\end{align}%
In words, $K_{2}$ acts in the conformal diagram and the first $2p$ rotation
planes as~$K_{1}$, permutes the last three (non-rotating) rotation planes
cyclically and rotates in one of them by $2\pi t/(3s)$. $K_{2}$~generates an
isometry group of order $4q=9s$, this group acts freely and properly
discontinuously, and the quotient is a geon. Spatial infinity is an order $%
2q $ quotient of $S^{4p+5}$, the lowest possible value of $q$ now being~$9$.

In both examples the geon has the $p$ rotational Killing vectors $\partial
_{\theta _{i}}+\partial _{\varphi _{i}}$, $i=1,\ldots ,p$, and in the second
example there is an additional rotational Killing vector $\partial _{\psi
_{1}}+\partial _{\psi _{2}}+\partial _{\psi _{3}}$. The angular momenta with
respect to all of these Killing vectors vanish. The $p$ independent
nonvanishing angular momenta are defined with respect to the Killing line
fields $\pm \left( \partial _{\theta _{i}}-\partial _{\varphi _{i}}\right) $%
, $i=1,\ldots ,p$, which can be defined as vector fields in a neighbourhood
of infinity but only as line fields on the whole geon. That a rotational
Killing line field can have nonvanishing angular momentum on a manifold with
one asymptopia was noted in the four-dimensional setting in~\cite%
{fried-mayer}.

Further examples arise by generalising the isometries in each $(\theta
_{i},\varphi _{i})$ pair and in the $(\psi _{1},\psi _{2},\psi _{3})$ triple
in non-cyclic ways that can be read off from the list of block-diagonal
isometries of the round 3-sphere and the round 5-sphere on pages 224-225 in~%
\cite{wolf}, and also by using isometries that cyclically permute larger
numbers of angles. For example, when there are $2p$ nonvanishing angular
momenta such that the first $p$ are equal and the remaining $p$ have the
opposite sign, we can arrange the isometries to permute the $2p$ rotation
planes cyclically. We shall not attempt to enumerate all the possibilities.
We suspect however that the quotients shown above have the largest possible
number of unequal angular momenta.

Inclusion of dilatonic fields and antisymmetric tensor fields can in
principle be discussed as in section~\ref{sec:static}, although few such
solutions are at present explicitly known \cite%
{cvetic-youm,llatas,cvetic-lu-pope1,cvetic-lu-pope2,%
madden-ross, chong-cvetic-lu-pope1,chong-cvetic-lu-pope2}.

\section{Einstein-$\mathrm{SU}(2)$ geon}

\label{sec:su2geon}

Since the initial discovery of spherically symmetric asymptotically flat
Einstein-$\mathrm{SU}(2)$ black holes~\cite{bizon,kuenzle-masood}, a large
variety of Einstein-Yang-Mills black holes have been found. A~fairly recent
survey is given in~\cite{volkov-galtsov-rev}. The nonabelian gauge freedom
makes investigation of geon quotients substantially more complicated than in
the abelian case, especally as the known solutions tend to be given in
gauges in which the spacetime symmetries are not manifest. We address here
only one example, 
the four-dimensional spherically symmetric Einstein-$\mathrm{SU}(2)$ 
hole in which the cosmological constant may be negative or vanishing
\cite%
{bizon,kuenzle-masood,win-stability,bjoraker-hoso-small,bjoraker-hoso-big,win-sar-even,win-sar-odd}%
. We show that this hole has a geon variant with a trivial gauge bundle.

We follow the notation of~\cite{bjoraker-hoso-big}. The exterior metric is
written in Schwarzschild-like coordinates as 
\begin{equation}
\mathrm{d} s^2 = - \frac{H}{p^2} \, \mathrm{d} t^2 + \frac{\mathrm{d} r^2}{H}
+ r^2 \, \mathrm{d}\Omega^2 \ \ ,
\end{equation}
where $H$ and $p$ are functions of~$r$, $\mathrm{d}\Omega^2 = \mathrm{d}%
\theta^2 + \sin^2 \! \theta \, \mathrm{d}\varphi^2$ is the metric on the
unit two-sphere and $(\theta,\varphi)$ are the usual angle coordinates.
There is a nondegenerate horizon at $r=r_h>0$, the exterior is at $r_h < r <
\infty$, and the metric functions have the near-horizon expansions 
\begin{subequations}
\begin{align}
H &= \sum_{k=1}^{\infty} h_k {(r-r_h)}^k \ \ , \\
p &= 1 + \sum_{k=1}^{\infty} p_k {(r-r_h)}^k \ \ ,
\end{align}
where $h_1>0$. Transforming the gauge field configuration from the singular
gauge used in \cite{bjoraker-hoso-big} to a regular gauge, by the matrix (9)
in \cite{bjoraker-hoso-big} with $\Omega =\pi /2$, we obtain the gauge
potential 
\end{subequations}
\begin{equation}
\boldsymbol{A}^{(0)}= \frac{1}{2e} \left[ (\vec{n} \cdot \vec{\tau}) u \, 
\mathrm{d} t + (w-1) \epsilon _{ijk}\tau^{i} n^{j} \mathrm{d} n^{k} \right]
\ \ ,  \label{eq:Anought}
\end{equation}
where $\tau ^{i}$ are the Pauli matrices as in~\cite{bjoraker-hoso-big}, $%
\vec{n}:=(\sin \theta \cos \varphi ,\sin \theta \sin \varphi ,\cos \theta )$%
, and $u$ and $w$ are functions of $r$ with the near-horizon expansions 
\begin{subequations}
\begin{align}
u &= \sum_{k=1}^{\infty} u_k {(r-r_h)}^k \ \ , \\
w &= \sum_{k=0}^{\infty} w_k {(r-r_h)}^k \ \ .
\end{align}

We define the Kruskal coordinates in the exterior by $U = - \exp \bigl[%
-\tfrac12 h_1(t-\rho)\bigr]$ and $V = \exp \bigl[\tfrac12 h_1(t+\rho)\bigr]$%
, where $\mathrm{d}\rho / \mathrm{d} r = p/H$ and we choose the additive
constant in $\rho$ so that $\rho = h_1^{-1} \ln\bigl[(r-r_h)\bigr] + 
\mathcal{O}\bigl( (r-r_0) \bigr)$. The horizon is at $UV\to0$, $r$ is a
function of $UV$ with the Taylor expansion $r = r_h - UV + \mathcal{O}\bigl( %
(UV)^2 \bigr)$, and the metric takes the form 
\end{subequations}
\begin{equation}
\mathrm{d} s^2 = - f \, \mathrm{d} U \mathrm{d} V + r^2 \mathrm{d}\Omega^2 \
\ ,  \label{eq:BH-krusalmetric}
\end{equation}
where $f$ is a function of $UV$ with the Taylor expansion $f = 4 h_1^{-1} + 
\mathcal{O}( UV )$. The gauge potential (\ref{eq:Anought}) takes the form 
\begin{equation}
\boldsymbol{A}^{(0)}= \frac{1}{2e} \left[ (\vec{n} \cdot \vec{\tau}) g \, (V 
\mathrm{d} U - U \mathrm{d} V) + (w-1) \epsilon _{ijk}\tau^{i} n^{j} \mathrm{%
d} n^{k} \right] \ \ ,  \label{eq:AnoughtKruskal}
\end{equation}
where $g$ is a function of $UV$ with the Taylor expansion $g = u_1 h_1^{-1}
+ \mathcal{O}( UV )$. The solution given by (\ref{eq:BH-krusalmetric}) and (%
\ref{eq:AnoughtKruskal}) is clearly extendible across $UV=0$ into a
two-exterior black hole spacetime $(\mathcal{M}, \mathrm{d} s^2)$ in the
standard fashion. As $\boldsymbol{A}^{(0)}$ (\ref{eq:AnoughtKruskal}) 
then becomes a globally-defined $\mathfrak{su}(2)$-valued
1-form on~$\mathcal{M}$, the gauge bundle over $\mathcal{M}$ is
trivial.

Now, the only involutive freely-acting isometry on the two-sphere is
the antipodal map $P:(\theta ,\varphi )\mapsto (\pi -\theta ,\varphi
+\pi )$, or $\vec{n}\mapsto -\vec{n}$. With this~$P$, the isometry 
$J$ (\ref{eq:J-action}) plainly leaves 
$\boldsymbol{A}^{(0)}$ (\ref{eq:AnoughtKruskal}) invariant. 
The quotient of $(\mathcal{M}, \mathrm{d} s^2)$ by 
$\bigl\{\mathrm{Id}_{\mathcal{M}},J\bigr\}$ 
is therefore a geon with a trivial gauge bundle. 
The ranges of the magnetic and
electric charges in the geon are exactly as in the two-exterior hole.

The difference from the 
spherically symmetric $\mathrm{U}(1)$ geons, 
in which both the electric and magnetic field strengths needed 
to be twisted, arises from
the nontrivial $\mathfrak{su}(2)$-valuedness of the gauge field. It can be
verified that the `electric' and `magnetic' parts of the field strength
computed from $\boldsymbol{A}^{(0)}$ (\ref{eq:AnoughtKruskal}) are both in a
hedgehog-like $\mathfrak{su}(2)$ configuration over the two-sphere.

\section{Conclusions and discussion}

\label{sec:discussion}

In this paper we have constructed new families of geon-type black
holes in $D\geq 4$ dimensions for Einstein's theory coupled to gauge
fields, by taking $\mathbb{Z}_{2}$ quotients of eternal black hole
solutions that have a nondegenerate Killing horizon. In the static
vacuum case, the existence of the quotient was equivalent to the
existence of suitable isometries on the spatial sections, and
antisymmetric tensor fields with a pure $\boldsymbol{F}^{2}$ action
could then always be included by 
taking the field strength to be a section of
a suitable, and in most cases nontrivial, bundle. With angular momentum,
we constructed rotating geons as quotients of the Myers-Perry(-AdS)
solution when $D$ is odd and not equal to~$7$. These rotating geons
are asymptotically flat (anti-de~Sitter, respectively), but the 
spatial infinity is a non-trivial quotient of the usual sphere. For
other $D$ we showed that rotating geon quotients of the
Myers-Perry(-AdS) solution, if they exist at all, cannot be
continuously deformed to the static case. With a negative cosmological
constant, we constructed geons with angular momenta on a torus at the
infinity. Finally, we showed that the $D=4$ spherically symmetric
$\mathrm{SU}(2)$ black hole admits a geon quotient with a trivial
gauge bundle.

In the Gibbons-Wiltshire geons of section~\ref{sec:static}, we
analysed the abelian gauge field configuration in terms of
the field strength, rather than in terms of the gauge potential. In
the electric geons and in the $D=4$ magnetic geons, the gauge field
can be interpreted as a connection in a principal $\mathrm{O}(2)$
bundle. For the electric geons and for the magnetic $D=4$ geons
with $\overline{\lambda}\le0$, this follows by first 
interpreting the gauge
field on the universal covering space as a connection in the trivial
$\mathrm{O}(2)$ bundle and then using the disconnected component of
$\mathrm{O}(2)$ to twist the bundle on taking the quotient. For the
$D=4$ magnetic geon with $\overline{\lambda}=1$, a similar argument is
not available since the $\mathrm{U}(1)
\simeq \mathrm{SO}(2)$ bundle on
the universal covering space is already nontrivial, but the existence
of the $\mathrm{O}(2)$ bundle can be verified
directly.\footnote{This is essentially the observation that the
$\mathrm{SO}(2)$ monopole on $S^2$ with an even monopole number can be 
quotiented into an $\mathrm{O}(2)$ monopole on~$\RPtwo$.}
For the $D>4$ magnetic geons, the gauge potential issue would need to
be addressed in terms of gerbes~\cite{mackaay-picken}.

All our solutions have a high degree of symmetry. For the $D=4$ solutions
with an abelian gauge field, this symmetry is dictated by the uniqueness
theorems for stationary black holes~\cite{heusler-book}. The uniqueness
theorems are known not to have a straightforward generalisation to $D>4$ %
\cite{emparan-reall-ring,elvang-etall-susyring}, and there are suggestions
that the uniqueness violation could be severe~\cite{reall-susy-unique}. With
a nonabelian gauge field, black holes with less symmetry are known to occur
even in the static $D=4$ context \cite%
{klei-kunz-axialprl,klei-kunz-axialprd,radu-winstanley}. One would like to
understand to what extent our geon quotients generalise to black holes with
less symmetry. One would also like to understand at a more general level
what kind of hair a geon may have. For example, despite the similarities
between the spherically symmetric $\mathrm{SU}(2)$ black hole and the
spherically symmetric skyrmion black hole of \cite%
{luckock-moss,luckock-conf,droz-heusler-strau-new}, the skyrmion field
configuration is not invariant under the isometry that yielded the $\mathrm{%
SU}(2)$ geon in section~\ref{sec:su2geon}. What is the situation for, say,
the spherically symmetric $\mathrm{SU}(n)$ black holes with $n>2$ \cite%
{kuenzle-sun}?

{}From the viewpoint of string theory, it would be of interest to explore
geon versions of black holes that are solutions to supergravity limits of
M-theory. String theoretic dualities have provided a microscopic
state-counting explanation for the entropy of certain families of such black
holes~\cite{peet-tasi}: can D-brane states corresponding to geons be
identified and distinguished from those corresponding to other black holes?
If yes, do such states successfully account for the physical entropy of the
geon?

\section*{Acknowledgements}

We thank John Barrett, John Friedman, Nico Giulini, Eli Hawkins, Viqar
Husain, Hans-Peter K\"{u}nzle, Todd Oliynyk, Simon Ross, David Wiltshire,
Elizabeth Winstanley and Jacek Wi\'{s}niewski for helpful discussions. J.~L.
was supported in part by the Engineering and Physical Sciences Research
Council. R.~B.~M. was supported in part by the Natural Sciences and
Engineering Research Council of Canada. D.~M. was supported in part by NSF
grant PHY03-54978 and by funds from the University of California.


\newpage

\begin{thebibliography}{999}
\bibitem{wheeler-geon} 
J.~A. Wheeler, 
``Geons'', 
{\it Phys.\ Rev.\ \bf 97}, 511 
(1955).

\bibitem{brill-wheeler-geon} 
D.~R. Brill 
and 
J.~A. Wheeler, 
``Interaction of
neutrinos and gravitational fields'', 
{\it Rev.\ Mod.\ Phys.\ \bf 29}, 465 
(1957).

\bibitem{ernst-geon-var} 
F.~J. Ernst, Jr., 
``Variational calculations in
geon theory'', 
{\it Phys.\ Rev.\ \bf 105}, 1662 
(1957).

\bibitem{ernst-geon} 
F.~J. Ernst, Jr., 
``Linear and toroidal geons'', 
{\it Phys.\ Rev.\ \bf 105}, 1665 
(1957).

\bibitem{MW-geondata} 
C.~W. Misner 
and 
J.~A. Wheeler, 
``Classical physics as
geometry: Gravitation, electromagnetism, unquantized charge, and mass as
properties of curved empty space'' 
{\it Ann.\ Phys.\ (NY) \bf 2}, 525 
(1957). 
Reprinted in: J.~A. Wheeler, \textit{Geometrodynamics\/} (Academic,
New York, 1962).

\bibitem{brill-hartle-geon} 
D.~R. Brill 
and 
J.~B. Hartle, 
``Method of the
self-consistent field in general relativity and its application to the
gravitational geon'', 
{\it Phys.\ Rev.\ \bf 135}, B271 
(1964).

\bibitem{melvin-pl} 
M.~A. Melvin, 
``Pure magnetic and electric geons'', 
{\it Phys.\ Lett.\ \bf 8}, 65 
(1963).

\bibitem{melvin-prd} 
M.~A. Melvin, 
``Dynamics of cylindrical electromagnetic
universes'', 
{\it Phys.\ Rev.\ \bf 139}, B225 
(1965).

\bibitem{thorne-Cen} 
K.~S. Thorne, 
``Energy of infinitely long,
cylindrically symmetric systems in general relativity'', 
{\it Phys.\ Rev.\ \bf 138}, B251 
(1965).

\bibitem{thorne-melvinstab} 
K.~S. Thorne, 
``Absolute Stability of Melvin's
Magnetic Universe'', 
{\it Phys.\ Rev.\ \bf 139}, B244 
(1965).

\bibitem{gundlach-rev} 
C.~Gundlach, 
``Critical phenomena in gravitational
collapse'', 
{\it Phys.\ Reports \bf 376}, 339 
(2003). 
$\langle$\texttt{arXiv:gr-qc/0210101}$\rangle$

\bibitem{sorkin-topogeon}
R.~D. Sorkin, ``Introduction to topological geons'', in:
\textit{Topological properties and global structure of space-time\/},
Proceedings of the NATO Advanced Study Institute on on Topological
Properties and Global Structure of Space-Time, Erice, Italy, May
12-22, 1985, 
edited by 
P.~G. Bergmann and V. De~Sabbata
(Plenum, 1986), pp.\ 249--270. 

\bibitem{FriedSchWi} 
J.~L. Friedman, 
K.~Schleich 
and 
D.~M. Witt,
``Topological censorship'', 
{\it Phys.\ Rev.\ Lett.\ \bf 71}, 1486 (1993); 
Erratum, 
{\it Phys.\ Rev.\ Lett.\ \bf 75}, 1872 (1995). 
\lanln{gr-qc/9305017}

\bibitem{giulini-thesis} 
D.~Giulini, 
``3-manifolds in canonical quantum
gravity'', 
Ph.D. Thesis, 
University of Cambridge (1990).

\bibitem{giulini-multiRP3s} 
D.~Giulini, 
``Two-body interaction energies in
classical general relativity'', 
in: \textit{Relativistic Astrophysics and
Cosmology\/}, Proceedings of the Tenth Seminar, Potsdam, October 21-26 1991,
edited by S.~Gottl\"ober, J.~P. M\"ucket and V.~M\"uller (World Scientific,
Singapore, 1992), pp.\ 333--338.

\bibitem{szekeres} 
G.~Szekeres, 
``On the singularities of a Riemannian
manifold'', 
{\it Publ.\ Mat.\ Debrecen \bf 7}, 285 
(1960).

\bibitem{chamblin-gibbons} 
A.~Chamblin 
and 
G.~W. Gibbons, 
``Nucleating black
holes via nonorientable instantons'', 
{\it Phys.\ Rev.\ \rm D \bf 55}, 2177 
(1997). 
$\langle$\texttt{arXiv:gr-qc/9607079}$\rangle$

\bibitem{louko-marolf-rp3} 
J.~Louko 
and 
D.~Marolf, 
``Inextendible
Schwarzschild black hole with a single exterior: How thermal is the Hawking
radiation?'' 
{\it Phys.\ Rev.\ \rm D \bf 58}, 024007 (1998). 
\lanln{gr-qc/9802068}

\bibitem{louko-whiting} 
J.~Louko 
and 
B.~F. Whiting, 
``Hamiltonian
thermodynamics of the Schwarzschild black hole'', 
{\it Phys.\ Rev.\ \rm D \bf 51}, 5583 
(1995). 
$\langle$\texttt{arXiv:gr-qc/9411017}$\rangle$

\bibitem{friedman-louko-winters} 
J.~L. Friedman, 
J.~Louko 
and 
S.~N. Winters-Hilt, 
``Reduced phase space formalism for spherically symmetric
geometry with a massive dust shell'', 
{\it Phys.\ Rev.\ \rm D \bf 56}, 7674 
(1997). 
$\langle$\texttt{arXiv:gr-qc/9706051}$\rangle$

\bibitem{langlois-rp3} 
P.~Langlois, 
``Hawking radiation for Dirac spinors on
the $\mathbb{RP}^3$ geon'', 
{\it Phys.\ Rev.\ \rm D \bf 70}, 104008 (2004).
$\langle$\texttt{arXiv:gr-qc/0403011}$\rangle$

\bibitem{louko-marolf} J.~Louko and D.~Marolf, 
``Single-exterior black holes and the AdS-CFT conjecture'', 
{\it Phys.\ Rev.\ \rm D \bf 59},
066002 (1999). 
$\langle$\texttt{arXiv:hep-th/9808081}$\rangle$

\bibitem{lou-ma-ross} 
J.~Louko, 
D.~Marolf 
and 
S.~F. Ross, 
``Geodesic
propagators and black hole holography'', 
{\it Phys.\ Rev.\ \rm D \bf 62}, 044041 (2000). 
$\langle$\texttt{arXiv:hep-th/0002111}$\rangle$

\bibitem{malda-eternal} 
J.~M. Maldacena, 
``Eternal black holes in
anti-de~Sitter'', 
{\it JHEP\/ \bf 0304}, 021 (2003).
\lanln{hep-th/0106112}

\bibitem{aminneborg-bengtsson-holst} 
S.~{\AA}minneborg, 
I.~Bengtsson 
and
S.~Holst, 
``A spinning anti-de~Sitter wormhole'', 
{\it Class.\ Quantum Grav.\ \bf 16}, 363 
(1999). 
$\langle$\texttt{arXiv:gr-qc/9805028}$\rangle$

\bibitem{brill-samos} 
D.~Brill, 
``Black holes and wormholes in 2+1
dimensions'', 
in: 
\textit{Mathematical and Quantum Aspects of Relativity and
Cosmology\/}, Proceedings of 2nd Samos Meeting on Cosmology, Geometry and
Relativity Karlovasi, Greece, 31 August - 4 September 1998 (Lecture Notes in
Physics, Vol.\ 537), edited by S.~Cotsakis and G.~W. Gibbons (Springer,
Berlin, 2000), pp.\ 143--179. 
$\langle$\texttt{arXiv:gr-qc/9904083}$\rangle$

\bibitem{brill-weimar} 
D.~Brill, 
``(2+1)-dimensional black holes with
momentum and angular momentum'', 
{\it Annalen Phys.\ \rm (Leipzig) \bf 9}, 217 
(2000). 
$\langle$\texttt{arXiv:gr-qc/9912079}$\rangle$

\bibitem{myers-perry} 
R.~C. Myers 
and 
M.~J. Perry, 
``Black holes in
higher-dimensional spacetimes'', 
{\it Ann.\ Phys.\ \rm (N.Y.) \bf 172}, 304 
(1986).

\bibitem{hawking-hunter-tr} 
S.~W. Hawking, 
C.~J. Hunter 
and 
M.~M. Taylor-Robinson, 
``Rotation and the AdS/CFT correspondence'', 
{\it Phys.\ Rev.\ \rm D \bf 59}, 064005 (1999). 
\lanln{hep-th/9811056}

\bibitem{gibbons-lu-page-pope1} 
G.~W. Gibbons, 
H.~L\"u, 
D.~N. Page 
and 
C.~N. Pope, 
``Rotating black holes in higher dimensions with a cosmological constant'', 
{\it Phys.\ Rev.\ Lett.\ \bf 93}, 171102
(2004). 
$\langle$\texttt{arXiv:hep-th/0409155}$\rangle$

\bibitem{gibbons-lu-page-pope2} 
G.~W. Gibbons, 
H.~L\"u, 
D.~N. Page 
and 
C.~N. Pope, 
``The general Kerr-de~Sitter metrics in all dimensions'', 
{\it J. Geom.\ Phys.\ \bf 53}, 49
(2005). 
$\langle$\texttt{arXiv:hep-th/0404008}$\rangle$

\bibitem{gibbons-perry-pope} 
G.~W. Gibbons, 
M.~J. Perry 
and 
C.~N. Pope,
``The first law of thermodynamics for Kerr-Anti-de~Sitter black
holes''. 
\lanln{hep-th/0408217}

\bibitem{bizon} 
P.~Bizon, 
``Colored black holes'', 
{\it Phys.\ Rev.\ Lett.\ \bf 64}, 2844 
(1990).

\bibitem{kuenzle-masood} 
H.~P. K\"unzle 
and 
A.~K.~M. Masood-ul-Alam,
``Spherically symmetric static $\mathrm{SU}(2)$ Einstein-Yang-Mills
fields'', 
{\it J.\ Math.\ Phys.\ \bf 31}, 928 
(1990).

\bibitem{win-stability} 
E.~Winstanley, 
``Existence of stable hairy black
holes in SU(2) Einstein-Yang-Mills theory with a negative cosmological
constant'', 
{\it Class.\ Quantum Grav.\ \bf 16}, 1963 
(1999). 
$\langle$\texttt{arXiv:gr-qc/9812064}$\rangle$

\bibitem{bjoraker-hoso-small} 
J.~Bjoraker 
and 
Y.~Hosotani, 
``Stable monopole
and dyon solutions in the Einstein-Yang-Mills theory in asymptotically
anti-de~Sitter space'', 
{\it Phys.\ Rev.\ Lett.\ \bf 84}, 1853
(2000). 
$\langle$\texttt{arXiv:gr-qc/9906091}$\rangle$

\bibitem{bjoraker-hoso-big} 
J.~Bjoraker 
and 
Y.~Hosotani, 
``Monopoles, dyons
and black holes in the four-dimensional Einstein-Yang-Mills theory'', 
{\it Phys.\ Rev.\ \rm D \bf 62}, 043513 (2000). 
\lanln{hep-th/0002098}

\bibitem{win-sar-even} 
E.~Winstanley 
and 
O.~Sarbach, ``On the linear
stability of solitons and hairy black holes with a negative cosmological
constant: The even parity sector'', 
{\it Class.\ Quantum Grav.\ \bf 19}, 689 
(2002). 
$\langle$\texttt{arXiv:gr-qc/0111039}$\rangle$

\bibitem{win-sar-odd} 
E.~Winstanley 
and 
O.~Sarbach, 
``On the linear
stability of solitons and hairy black holes with a negative cosmological
constant: The odd parity sector'', 
{\it Class.\ Quantum Grav.\ \bf 18}, 2125 
(2001). 
$\langle$\texttt{arXiv:gr-qc/0102033}$\rangle$

\bibitem{GibbWilt} 
G.~W. Gibbons 
and 
D.~L. Wiltshire, 
``Space-time as a
membrane in higher dimensions'', 
{\it Nucl.\ Phys.\ \bf B287}, 717
(1987). 
$\langle$\texttt{arXiv:hep-th/0109093}$\rangle$

\bibitem{lemos1} 
J.~P.~S. Lemos, 
``Two-dimensional black holes and planar
general relativity'', 
{\it Class.\ Quantum Grav.\ \bf 12}, 1081 
(1995). 
$\langle$\texttt{arXiv:gr-qc/9407024}$\rangle$

\bibitem{lemos2} 
J.~P.~S. Lemos, 
``Cylindrical black hole in general
relativity'', 
{\it Phys.\ Lett.\ \bf B353}, 46 
(1995). 
$\langle$\texttt{arXiv:gr-qc/9404041}$\rangle$

\bibitem{huang-liang} 
C.~Huang 
and 
C.~Liang, 
``A torus-like black hole'', 
{\it Phys.\ Lett.\ \bf A201}, 27 
(1995).

\bibitem{lemos-zanchin} 
J.~P.~S. Lemos 
and 
V.~T. Zanchin, 
``Rotating charged
black string and three-dimensional black holes'', 
{\it Phys.\ Rev.\ \rm D \bf 54}, 3840 
(1996). 
$\langle$\texttt{arXiv:hep-th/9511188}$\rangle$

\bibitem{mann-pair} 
R.~B. Mann, 
``Pair production of topological
anti-de~Sitter black holes'', 
{\it Class.\ Quantum Grav.\ \bf 14},
L109 
(1997). 
$\langle$\texttt{arXiv:gr-qc/9607071}$\rangle$

\bibitem{cai-zhang} 
R.~Cai 
and 
Y.~Zhang, 
``Black plane solutions in four
dimensional spacetimes'', 
{\it Phys.\ Rev.\ \rm D \bf 54}, 4891 
(1996). 
$\langle$\texttt{arXiv:gr-qc/9609065}$\rangle$

\bibitem{smith-mann} 
W.~L. Smith 
and 
R.~B. Mann, 
``Formation of topological
black holes from gravitational collapse'', 
{\it Phys.\ Rev.\ \rm D \bf 56}, 4942 
(1977). 
$\langle$\texttt{arXiv:gr-qc/9703007}$\rangle$

\bibitem{banados} 
M.~Ba\~{n}ados, 
``Constant curvature black holes'', 
{\it Phys.\ Rev.\ \rm D \bf 57}, 1068 (1998). 
\lanln{gr-qc/9703040}

\bibitem{vanzo} 
L.~Vanzo, 
``Black holes with unusual topology'', 
{\it Phys.\ Rev.\ \rm D \bf 56}, 6475 
(1997). 
$\langle$\texttt{arXiv:gr-qc/9705004}$\rangle$

\bibitem{mann-negmass} 
R.~Mann, 
``Black holes of negative mass'', 
{\it Class.\ Quantum Grav.\ \bf 14}, 2927 
(1997). 
$\langle$\texttt{arXiv:gr-qc/9705007}$\rangle$

\bibitem{brill-louko-peldan} 
D.~R. Brill, 
J.~Louko 
and 
P.~Peld\'an,
``Thermodynamics of $(3+1)$-dimensional black holes with toroidal or higher
genus horizons'', 
{\it Phys.\ Rev.\ \rm D \bf 56}, 3600 
(1997). 
$\langle$\texttt{arXiv:gr-qc/9705012}$\rangle$

\bibitem{birmingham-topol} 
D.~Birmingham, 
``Topological black holes in
anti-de Sitter space'', 
{\it Class.\ Quantum Grav.\ \bf 16}, 1197
(1999). 
$\langle$\texttt{arXiv:hep-th/9808032}$\rangle$

\bibitem{emparan-johnson-myers} 
R.~Emparan, 
C.~V. Johnson 
and 
R.~C. Myers,
``Surface terms as counter\-terms in the AdS/CFT correspondence'', 
{\it Phys.\ Rev.\ \rm D \bf 60}, 104001 (1999). 
\lanln{hep-th/9903238}

\bibitem{emparan} 
R.~Emparan, 
``AdS/CFT duals of topological black holes and
the entropy of zero energy states'', 
{\it JHEP\/ \bf 9906}, 036
(1999). 
$\langle$\texttt{arXiv:hep-th/9906040}$\rangle$

\bibitem{GibbMaeda} 
G.~W. Gibbons 
and 
K.~Maeda, 
``Black holes and membranes
in higher-dimensional theories with dilaton fields'', 
{\it Nucl.\ Phys.\ \bf B298}, 741 (1988).

\bibitem{bott-tu} 
R.~Bott 
and 
L.~W. Tu, 
{\it Differential Forms in Algebraic Topology\/} 
(Springer, New York, 1982).

\bibitem{wald-qft} 
R.~M. Wald, 
\textit{Quantum Field Theory in Curved
Spacetime and Black Hole Thermodynamics\/} (The University of Chicago Press,
Chicago, 1994).

\bibitem{haw-ell} 
S.~W. Hawking 
and 
G.~F.~R. Ellis, 
\textit{The Large Scale
Structure of Space-Time\/} (Cambridge University Press, Cambridge, 1973).

\bibitem{lake-rnds} 
K.~Lake, 
``Reissner-Nordstr\"om-de Sitter metric, the
third law, and cosmic censorship'', 
{\it Phys.\ Rev.\ \rm D \bf 19}, 421 
(1979).

\bibitem{strobl2} 
T.~Kl\"{o}sch 
and 
T.~Strobl, 
``Classical and quantum
gravity in 1+1 dimensions. II: The universal coverings'', 
{\it Class.\ Quantum Grav.\ \bf 13}, 2395 (1996). 
$\langle$\texttt{arXiv:gr-qc/9511081}$\rangle$

\bibitem{Fid-etal} 
L.~Fidkowski, 
V.~Hubeny, 
M.~Kleban 
and 
S.~Shenker, 
``The black hole singularity in AdS/CFT'', 
{\it JHEP\/ \bf 0402}, 014
(2004). 
$\langle$\texttt{arXiv:hep-th/0306170}$\rangle$

\bibitem{gal-sch-witt}
G.~Galloway, 
K.~Schleich
and 
D.~M. Witt, 
``Topological censorship and higher genus black holes'', 
{\it Phys.\ Rev.\ \rm D \bf 60}, 104039 
(1999). 
\lanln{gr-qc/9902061}

\bibitem{gal-sch-witt-wool}
G.~Galloway, 
K.~Schleich, 
D.~M. Witt 
and 
E.~Woolgar, 
``The AdS/CFT correspondence conjecture and topological censorship'', 
{\it Phys.\ Lett.\ \bf B505}, 255
(2001). 
\lanln{hep-th/9912119}

\bibitem{sorkin-relation} 
R.~Sorkin, 
``On the relation between charge and
topology'', 
{\it J.\ Phys.\ \rm A \bf 10}, 717 
(1977).

\bibitem{sorkin-multiply} 
R.~Sorkin, 
``The quantum electromagnetic field in
multiply connected space'', 
{\it J.\ Phys.\ \rm A \bf 12}, 403 
(1979).

\bibitem{fried-mayer} 
J.~L. Friedman 
and 
S.~Mayer, 
``Vacuum handles carrying
angular momentum; electrovac handles carrying net charge'', 
{\it J.\ Math.\ Phys.\ \bf 23}, 109 (1982).

\bibitem{hatcher} 
A.~Hatcher, 
\textit{Algebraic Topology\/} 
(Cambridge University Press, Cambridge, 2002), 
Proposition 1.40 and Exercise 1.3.24.

\bibitem{wolf} 
J.~A. Wolf, 
{\it Spaces of Constant Curvature\/}, 5th
edition (Publish or Perish, Wilmington, 1984).

\bibitem{horo-maro} 
G.~T. Horowitz 
and 
D.~Marolf, 
``A new approach to string
cosmology'', 
{\it JHEP\/ \bf 9807}, 014 (1998). 
\lanln{hep-th/9805207}

\bibitem{schiffer-spencer} 
M.~Schiffer 
and 
D.~C. Spencer, 
\textit{Functionals of finite Riemann surfaces\/} (Princeton University Press,
Princeton, New Jersey, 1954).

\bibitem{alling-greenleaf} 
N.~L. Alling 
and 
N.~Greenleaf, 
\textit{Foundations of the theory of Klein surfaces\/}, 
Lecture Notes in Mathematics
Vol.~219 (Springer, Berlin, 1971).

\bibitem{duff-nilsson-pope} 
M.~J. Duff, 
B.~E.~W. Nilsson 
and 
C.~N. Pope,
``Kaluza-Klein supergravity'', 
{\it Phys.\ Reports \bf 130}, 1 
(1986).

\bibitem{ChanHorneMann} 
K.~C.~K. Chan, 
J.~Horne 
and 
R.~B. Mann, 
``Charged
Dilaton black holes with unusual asymptotics'', 
{\it Nucl.\ Phys.\ \bf B447}, 441 
(1995). 
$\langle$\texttt{arXiv:gr-qc/9502042}$\rangle$

\bibitem{louko-schleich} 
J.~Louko 
and 
K.~Schleich, 
``The exponential law:
Monopole detectors, Bogolubov transformations, and the thermal nature of the
Euclidean vacuum in $\mathbb{RP}^3$ de~Sitter spacetime'', 
{\it Class.\ Quantum Grav. \bf 16}, 2005 
(1999). 
$\langle$\texttt{arXiv:gr-qc/9812056}$\rangle$

\bibitem{schleich-witt-rp3} 
K.~Schleich 
and 
D.~M. Witt, 
``The generalized
Hartle-Hawking initial state: Quantum field theory on Einstein conifolds'', 
{\it Phys.\ Rev.\ \rm D \bf 60}, 064013 (1999). 
\lanln{gr-qc/9903062}

\bibitem{McInnes-schwds} 
B.~McInnes, 
``de~Sitter and Schwarzschild-de~Sitter
according to Schwarzschild and de~Sitter'', 
{\it JHEP\/ \bf 0309}, 009 (2003). 
$\langle$\texttt{arXiv:hep-th/0308022}$\rangle$

\bibitem{gh-deS} 
G.~W. Gibbons 
and 
S.~W. Hawking, 
``Cosmological event
horizons, thermodynamics, and particle creation'', 
{\it Phys.\ Rev.\ \rm D \bf 15}, 2738 
(1977).

\bibitem{strobl3} 
T.~Kl\"{o}sch 
and 
T.~Strobl, 
``Classical and quantum
gravity in 1+1 dimensions. III: Solutions with arbitrary topology'', 
{\it Class.\ Quantum Grav.\ \bf 14}, 1689 
(1997). 
$\langle$\texttt{arXiv:hep-th/9607226}$\rangle$

\bibitem{witten-dscft} 
E.~Witten, 
``Quantum gravity in de~Sitter space'',
in: \textit{New Fields and Strings in Subnuclear Physics\/}, edited by
A.~Zichichi (Singapore, World Scientific, 2003) 
[{\it Int.\ J. Mod.\ Phys.\ \rm A \bf 18}, Supplement (2003)]. 
\lanln{hep-th/0106109}

\bibitem{strominger-dscft} 
A.~Strominger, 
``The dS/CFT correspondence'', 
{\it JHEP\/ \bf 0110}, 034 (2001). 
\lanln{hep-th/0106113}

\bibitem{aminneborg} 
S.~{\AA }minneborg, 
I.~Bengtsson, 
S.~Holst 
and 
P.~Peld\'{a}n, 
``Making anti-de Sitter black holes'', 
{\it Class.\ Quantum Grav.\ \bf 13}, 2707 (1996). 
\lanln{gr-qc/9604005}

\bibitem{holst} 
S.~Holst 
and 
P.~Peld\'{a}n, 
``Black holes and causal
structure in anti-de Sitter isometric spacetimes'', 
{\it Class.\ Quantum Grav.\ \bf 14}, 3433 (1997). 
\lanln{gr-qc/9705067}

\bibitem{CreightonMann} 
J.~D.~E. Creighton 
and 
R.~B. Mann, 
``Entropy of
constant curvature black holes in general relativity'', 
{\it Phys.\ Rev.\ \rm D \bf 58}, 024013 (1998). 
\lanln{gr-qc/9710042}

\bibitem{gomberoff} 
M.~Ba\~{n}ados, 
A.~Gomberoff 
and 
C.~Mart\'{\i}nez,
``Anti-de Sitter space and black holes'', 
{\it Class.\ Quantum Grav.\ \bf 15}, 3575 (1998). 
$\langle$\texttt{arXiv:hep-th/9805087}$\rangle$

\bibitem{louko-wisniewski} 
J.~Louko 
and 
J.~Wi\'sniewski, 
``Einstein black
holes, free scalars and AdS/CFT correspondence'', 
{\it Phys.\ Rev.\ \rm D \bf 70} 084024
(2004). 
\lanln{hep-th/0406140}

\bibitem{awad} 
A.~M. Awad, 
``Higher-dimensional charged rotating solutions
in (A)dS spacetimes'', 
{\it Class.\ Quantum Grav.\ \bf 20}, 2827 
(2003). 
$\langle$\texttt{arXiv:hep-th/0209238}$\rangle$

\bibitem{cvetic-youm} 
M.~Cvetic 
and 
D.~Youm, 
``Near BPS saturated rotating
electrically charged black holes as string states'', 
{\it Nucl.\ Phys.\ \bf B477}, 449 
(1996). 
$\langle$\texttt{arXiv:hep-th/9605051}$\rangle$

\bibitem{llatas} 
P.~M. Llatas, 
``Electrically charged black holes for the
heterotic string compactified on a $(10-D)$ torus'', 
{\it Phys.\ Lett.\ \bf A397}, 63 
(1997). $\langle$\texttt{arXiv:hep-th/9605058}$\rangle$

\bibitem{cvetic-lu-pope1} 
M.~Cveti\v{c}, 
H.~L\"u 
and 
C.~N. Pope, 
``Charged
Kerr-de~Sitter black holes in five dimensions'', 
{\it Phys.\ Lett.\ \bf B598}, 273
(2004). 
\lanln{hep-th/0406196}

\bibitem{cvetic-lu-pope2} 
M.~Cveti\v{c}, 
H.~L\"u 
and 
C.~N. Pope, 
``Charged rotating black holes in five dimensional 
${\mathrm{U}(1)}^3$ gauged $\mathcal{N}=2$ supergravity'', 
{\it Phys.\ Rev.\ \rm D \bf 70} 081502
(2004). 
$\langle$\texttt{arXiv:hep-th/0407058}$\rangle$.

\bibitem{madden-ross} 
O.~Madden 
and 
S.~F. Ross, 
``On uniqueness of charged
Kerr-AdS black holes in five dimensions'', 
{\it Class.\ Quantum Grav.\ \bf 22}
515
(2005). 
$\langle$\texttt{arXiv:hep-th/0409188}$\rangle$.

\bibitem{chong-cvetic-lu-pope1} 
Z.-W. Chong, 
M.~Cveti\v{c}, 
H.~L\"u 
and 
C.~N. Pope, 
``Charged rotating black holes in four-dimensional 
gauged and ungauged supergravities'', 
\lanln{hep-th/0411045}

\bibitem{chong-cvetic-lu-pope2} 
Z.-W. Chong, 
M.~Cveti\v{c}, 
H.~L\"u 
and 
C.~N. Pope, 
``Non-extremal charged rotating black holes in seven-dimensional 
gauged supergravity'', 
\lanln{hep-th/0412094}

\bibitem{volkov-galtsov-rev} 
M.~S. Volkov 
and 
D.~V. Gal'tsov, 
``Gravitating
non-abelian solitons and black holes with Yang-Mills fields'', 
{\it Phys.\ Rept.\ \bf 319}, 1 
(1999). 
$\langle$\texttt{arXiv:hep-th/9810070}$\rangle$

\bibitem{mackaay-picken}
M~Mackaay 
and 
R.~Picken, 
``Holonomy and parallel transport for abelian gerbes'', 
{\it Adv.\ Math.\ \bf 170}, 287
(2002). 
\lanln{math.dg/0007053}

\bibitem{heusler-book} 
M.~Heusler, 
\textit{Black Hole Uniqueness Theorems\/}
(Cambridge University Press, Cambridge, 1996).

\bibitem{emparan-reall-ring} 
R.~Emparan 
and 
H.~S. Reall, 
``A rotating black ring solution in five dimensions'', 
{\it Phys.\ Rev.\ Lett.\ \bf 88}, 101101 (2002). 
$\langle$\texttt{arXiv:hep-th/0110260}$\rangle$

\bibitem{elvang-etall-susyring} 
H.~Elvang, 
R.~Emparan, 
D.~Mateos 
and 
H.~S. Reall, 
``A supersymmetric black ring'', 
{\it Phys.\ Rev.\ Lett.\ \bf 93}
211302
(2004). 
\lanln{arXiv:hep-th/0407065}. 

\bibitem{reall-susy-unique} 
H.~S. Reall, 
``Higher dimensional black holes
and supersymmetry'', 
{\it Phys.\ Rev.\ \rm D \bf 68}, 024024
(2003). 
$\langle$\texttt{arXiv:hep-th/0211290}$\rangle$

\bibitem{klei-kunz-axialprl} 
B.~Kleihaus 
and 
J.~Kunz, 
``Static black hole
solutions with axial symmetry'', 
{\it Phys.\ Rev.\ Lett.\ \bf 79},
1595 
(1997). 
$\langle$\texttt{arXiv:gr-qc/9704060}$\rangle$

\bibitem{klei-kunz-axialprd} 
B.~Kleihaus 
and 
J.~Kunz, 
``Static axially
symmetric Einstein-Yang-Mills dilaton solutions. II. Black hole solutions', 
{\it Phys.\ Rev.\ \rm D \bf 57}, 6138 
(1998). 
$\langle$\texttt{arXiv:gr-qc/9712086}$\rangle$

\bibitem{radu-winstanley} 
E.~Radu 
and 
E.~Winstanley, 
``Static axially
symmetric solutions of Einstein-Yang-Mills equations with a negative
cosmological constant: Black hole solutions'', 
{\it Phys.\ Rev.\ \rm D \bf 70}, 084023 (2004). 
\lanln{hep-th/0407248}

\bibitem{luckock-moss} 
H.~Luckock 
and 
I.~Moss, 
``Black holes have skyrmion
hair'', 
{\it Phys.\ Lett.\ \bf B176}, 341 
(1986).

\bibitem{luckock-conf} 
H.~Luckock, 
``Black hole skyrmions'', 
in: 
\textit{String theory, quantum cosmology and quantum gravity: Proceedings of the
Paris-Meudon Colloquium 22-26 September 1986\/} (World Scientific,
Singapore, 1987), pp.\ 454--464.

\bibitem{droz-heusler-strau-new} 
S.~Droz, 
M.~Heusler 
and 
N.~Straumann, 
``New black hole solutions with hair'', 
{\it Phys.\ Lett.\ \bf B268}, 371 
(1991).

\bibitem{kuenzle-sun} 
H.~P. K\"unzle, 
``Analysis of the static spherically
symmetric $\mathrm{SU}(N)$ Einstein Yang-Mills equations'', 
{\it Commun.\ Math.\ Phys.\ \bf 162}, 371 
(1994).

\bibitem{peet-tasi} 
A.~W. Peet, 
``TASI lectures on black holes in string theory,'' 
in: 
\textit{Strings, Branes, and Gravity: TASI 99: Boulder,
Colorado, 31 May - 25 June 1999}, edited by J.~Harvey, S.~Kachru, and
E.~Silverstein (World Scientific, Singapore, 2001), pp.\ 353--433. 
\lanln{hep-th/0008241}

\end{thebibliography}
\end{document}